\documentclass[12pt]{article}
\pdfoutput=1
\usepackage[dvipsnames]{xcolor}
\usepackage[utf8]{inputenc}
\usepackage[english]{babel}
\usepackage{amsmath}
\usepackage{hyperref}
\usepackage{breakurl}
\usepackage[T1]{fontenc}
\usepackage{multicol}
\usepackage{mathtools}
\usepackage{float}
\usepackage{amssymb}
\usepackage{xfrac}
\usepackage{doi}
\usepackage[margin=0.75in]{geometry}
\usepackage[maxbibnames=99,sorting=none]{biblatex}
\usepackage{amsthm}
\usepackage{ytableau}
\usepackage{braket}
\usepackage{csquotes}
\usepackage{titling}
\AtBeginBibliography{}
\newcommand{\be}{\begin{equation}}
\newcommand{\ee}{\end{equation}}
\DeclareRobustCommand{\rchi}{{\mathpalette\irchi\relax}}
\newcommand{\irchi}[2]{\raisebox{\depth}{$#1\chi$}}
\usepackage{caption}
\usepackage{graphicx}
\usepackage{subfiles}

\usepackage{abstract}
\usepackage[makeroom]{cancel}
\usepackage{tensor}
\usepackage{youngtab}

\numberwithin{equation}{section}
\newtheorem{cor}{Corollary}[section]
\newtheorem{theo}{Theorem}[section]
\newtheorem{prop}{Proposition}[section]
\newtheorem{defi}{Definition}[section]
\newtheorem{assu}{Assumption}[section]
\newtheorem{rem}{Remark}[section]
\newtheorem{lemma}{Lemma}[section]
\begin{document}

\date{\vspace{-5ex}}
\title{%
  \textbf{\textit{\large{Time Reversal Symmetry for Classical, Non-relativistic Quantum and Spin Systems in Presence of Magnetic Fields }}}\\
    \large{\vspace{+2ex}Davide Carbone}\\
    \small{\vspace{+0.1ex}\textit{Dipartimento di Scienze Matematiche, Politecnico di Torino, Corso Duca degli Abruzzi 24, 10129 Torino, Italy}}\\
    \small{\vspace{+0.1ex}\textit{and INFN, Sezione di Torino, Via P. Giuria 1, 10125 Torino, Italy \\
    \normalfont{davide.carbone@polito.it}}}\\
    \large{\vspace{+2ex}Paolo De Gregorio}\\
    \small{\vspace{+0.1ex}\textit{Dipartimento di Scienze Matematiche, Politecnico di Torino, Corso Duca degli Abruzzi 24, 10129 Torino, Italy\\
    \normalfont{paolo.degregorio@polito.it}}}\\
    \large{\vspace{+2ex}Lamberto Rondoni}\\
    \small{\vspace{+0.1ex}\textit{Dipartimento di Scienze Matematiche, Politecnico di Torino, Corso Duca degli Abruzzi 24, 10129 Torino, Italy}}\\
    \small{\vspace{+0.1ex}\textit{and INFN, Sezione di Torino, Via P. Giuria 1, 10125 Torino, Italy \\
    \normalfont{lamberto.rondoni@polito.it}}}}
%
\leftskip=0cm
\rightskip=0cm
\maketitle
\noindent\textbf{Corresponding author}: Davide Carbone, davide.carbone@polito.it\\
\noindent\rule{\textwidth}{1pt}
\section*{\large Abstract}
We extend to quantum mechanical systems results previously obtained for classical mechanical systems, concerning time reversibility in presence of a magnetic field. As in the classical case, results like the Onsager reciprocal relations and the so-called fluctuation theorems, are consequently obtained, without recourse to the Casimir modification.
The quantum systems treated here are non-relativistic, and are described by the Schr{\"o}dinger equation or the Pauli equation. In particular, we prove that the spin-field interaction does not break the time reversal invariance (TRI) of the dynamics, and that it does not require additional conditions for such a symmetry to hold, compared to the spinless cases. These results are relevant for experiments such as diffusion in solutions, thermoelectricity and spin charge transport. Indeed, no violation of the Onsager relations has been found in presence of a magnetic field, contrary to  
general expectations.
\newline
\newline
\textit{Key words}: Onsager relation, magnetic field, time reversal, spin, invariance, symmetry transformation
\newline
\noindent\rule{\textwidth}{1pt}
\section{\large Introduction}
\label{sec:Introduction}
The present paper develops one aspect of response theory in presence of magnetic fields, or rotating reference frames, which until recently has been treated as a separate chapter of statistical mechanics response theory. The reason was that magnetic fields and rotating frames break the standard time reversal invariance of both classical and quantum mechanics, which until recently had been exclusively used to
derive the Onsager Reciprocal Relations \cite{onsager1931reciprocal,onsager1931reciprocal2}; or the 
Fluctuation-Dissipation Theorem and the Green-Kubo relations Refs.\cite{callen1951irreversibility,kubo1957statistical,kubo1959some,kubo1966fluctuation}; and, more lately, of some variation of the Fluctuation Theorem \cite{barbier:2018b}. Onsager stated that:

\vskip 3pt
\noindent
{\em "the principle of dynamical reversibility does not apply when (external) magnetic fields or Coriolis forces are present, and the
reciprocal relations break down"} \cite{onsager1931reciprocal}.

\vskip 3pt
\noindent
It was then observed by Casimir \cite{Casimir} that reversibility can be intended in an extended sense, in which two systems in two opposite magnetic fields are considered, rather than a single system. In the words of the celebrated Landau and Lifshitz book
\cite{landau1980statistical}: 

\vskip 3pt
\noindent
{\em "The proof depends on the symmetry of the equations of motion with respect to time, and the formulation of this symmetry is somewhat altered for fluctuations in a uniformly rotating body and for bodies in an external magnetic field: in these cases the symmetry under time reversal holds only if the sign of the angular velocity of rotation Q or of the magnetic field H is simultaneously changed."}

\vskip 3pt
\noindent
While formally correct, this approach
which has been universally adopted till recently \cite{lax,sachs,de2013non,toda1991statistical}
substantially weakens the predictive nature of the Onsager relations and, analogously, of the Fluctuation-Dissipation Theorem and the Green-Kubo relations; and of the Fluctuation Theorem. 
On the other hand, we are not aware of any experimental violation of the Onsager relations in presence of a magnetic field.

In Ref.\cite{robnik1986false}, Robnik and Berry considered generalized time reversal symmetries,
as associated to canonical time reversal, 
and investigated the entailing spectral fluctuations of quantum systems whose classical motion is chaotic. 
Nevertheless, the
standard symmetry known as microreversibility
has continued to be considered necessary for the validity of Onsager relations, cf.\ Ref.\cite{jacquod2012onsager} for
transport coefficients for coupled electric, spin, thermoelectric and spin caloritronic effects.


\vskip 3pt
\noindent
In recent times, it was noted that a form of time reversal invariance identified in the nonequilibrium molecular dynamics of shearing fluids  \cite{j2007statistical}, is suitable to frame within single systems response theory the response in presence of magnetic fields or rotating frames \cite{rondoe}. A series of works followed, showing an ever larger set of symmetries that play the role of time reversal invariance in the response theory of systems in magnetic fields, or equivalently in rotating frames \cite{rondom,bonella2017time,rondoquantu,rondoFT1,rondoFT2,carbone2020necessary}. In this paper, we complete the set of time reversal symmetries that can be used in presence of magnetic fields, for both classical and quantum systems, including non-relativistic spin systems. This is interesting, because the wider this set, the greater the predictive power of the theory. In the first part of the paper, we summarize the main aspects of linear response theory, highlighting the role of time reversibility, and we concisely describe our results. The formal derivations leading to such results are given in the second part of the paper.

\subsection{Fluctuation-Dissipation Theorem and Linear Response}
To illustrate linear response theory, we follow Ref.\cite{toda1991statistical}, that treats in parallel the classical and the quantum cases. The reader can avoid this subsection if already familiar with this topic. Consider a large Hamiltonian particle system initially in a canonical equilibrium, which is perturbed by a time dependent term, $H_{ext}(\boldsymbol X,\boldsymbol P;t)=A(\boldsymbol X,\boldsymbol P)F(t)$, where $\mathbf{X}$ and $\mathbf{P}$ are the canonically conjugate variables. Denoting by $H$ the unperturbed Hamiltonian, the dynamics is then governed by the following time dependent Hamiltonian:
\begin{equation}
    H_t(\boldsymbol X,\boldsymbol P)=H(\boldsymbol X,\boldsymbol P)-A(\boldsymbol X,\boldsymbol P)F(t)
\end{equation}
Let round brackets
\begin{equation}
    (\cdot , \cdot) \equiv
    \{\cdot , \cdot\}\;\;\;\;\;C.M.\ \, ; \qquad
    (\cdot , \cdot)\equiv\frac{i}{\hslash}[\cdot , \cdot]\;\;\;\;\; Q.M.
    \label{roundbr}
\end{equation}
respectively denote the Poisson brackets in
classical mechanics, and the commutator multiplied by $i/\hbar$ in quantum mechanics. Notice that the information on the statistical ensemble (bosons, fermions, anyons and so on) is automatically included in this formalism.
Introduce the symbol $\rho$ to represent the phase space probability distribution in classical mechanics, or the density matrix in quantum mechanics. Its evolution is  given by:
\begin{equation}
\label{lio}
    \frac{\partial \rho}{\partial t}=i\mathcal{L}_t\rho \, , 
     ~~ \mbox{where } ~
     i\mathcal{L}_t = -( H_t , \cdot ) =
     i\mathcal{L}+i\mathcal{L}_{ext} =
     -( H , \cdot ) -( H_{ext} , \cdot )
\end{equation}
where the contribution $\cal L$ due to $H$ and ${\cal L}_{ext}$ due to $H_{ext}$ are evidenced.
Let 
\begin{equation}
    \label{canonic}
    \rho_e=C\exp\{-\beta H\}
\end{equation}
describe the initial unperturbed canonical equilibrium state at temperature $T$, with $\beta=1/k_{_B} T$, $k_{_B}$ being the Boltzmann constant.
Assuming that the perturbation $H_{ext}$ is small, one may truncate to first order in $H_{ext}$ the solution of Eq.\eqref{lio}, obtaining:
\begin{equation}
\label{linrho}
    \rho(t)=\rho_e + \Delta\rho(t)+
    (\mbox{higher order terms in }H_{ext}) \, ; ~~
     \Delta\rho(t)=\int_{0}^t dt' e^{i(t-t')\mathcal{L}}i\mathcal{L}_{ext}(t')\rho_e
\end{equation}
Neglecting the higher order terms, leads to the linear response formula for any observable $B$, which gives the variation in time of its mean, $\mathbb{E}_t[\Delta B]$, as:
\begin{equation}
\label{2.59}
    \mathbb{E}_t[\Delta B]=\int_{0}^t dt' \, F(t') \, \textbf{Tr} \, \rho_e \, (A(0), B(t-t'))
\end{equation}
where $B(t)=e^{-i\mathcal{L}t}B$ in a classical framework, and $B(t)=e^{iHt/\hslash}Be^{-iHt/\hslash}$ in the quantum case, and the brackets $(\cdot , \cdot )$ has been defined by Eq.\eqref{roundbr}.
Moreover, to unify the classical and quantum mechanical notations as in Ref.\cite{toda1991statistical}, by {\bf Tr} we denote phase space integration
in classical mechanics, and trace operator applied to all the objects to the right of the symbol in quantum mechanics. Introducing the \textit{response function}:
\begin{equation}
\label{2.60}
    \phi_{BA}(t)\equiv\textbf{Tr}\rho_e(A(0), B(t))
\end{equation}
we may then write:
\begin{equation}
\label{2.61}
    \mathbb{E}_t[\Delta B]=\int_{0}^t dt' F(t') \phi_{BA}(t-t')
\end{equation}
Most beneficially, the linear response rests on the unperturbed equilibrium state $\rho_e$.
Under common conditions, we can rewrite
Eq.\eqref{2.60} as:
\begin{equation}
\label{ corre}
    \phi_{BA}(t)=\textbf{Tr}(\rho_e,A(0)) B(t) \, , ~~ \mbox{where } ~ ( \rho_e , A(0)) = \beta\rho_e \dot{A} \, . 
\end{equation}
The response formula states that the variation in time of the average of the observable $B$
is determined by the time integral of the equilibrium correlation function of $B$ and the perturbing Hamiltonian. The second equality in Eq.\eqref{ corre} is due to the
fact that $\rho_e$ is canonical.
In the quantum case, it is convenient to use the identity:
\begin{equation}
\label{2.69}
    [A,e^{-\beta H}]=e^{-\beta H}\int_0^\beta e^{\lambda H}[H,A]e^{-\lambda H}d\lambda
\end{equation}
and the fact that the evolution of an observable $B$ is given by $B(t)=e^{iHt/\hslash}Be^{-iHt/\hslash}$ which, using \eqref{ corre} and \eqref{2.69}, leads to
\begin{equation}
    e^{\lambda H}[H,A]e^{-\lambda H}=e^{\lambda H}\dot{A}e^{-\lambda H}=\dot{A}(-i\hslash \lambda)
\end{equation}
Eventually, in the quantum mechanical case, we can also write:
\begin{equation}
\label{2.71}
    (\rho_e,A)=\int_0^\beta \rho_e \dot{A}(-i\hslash \lambda)d\lambda
\end{equation}
The fundamental object of our investigation is the {\em canonical correlator}:
\begin{equation}
    \label{2.72}
    \mathbb{E}(X;Y)\equiv\langle X;Y\rangle=\frac{1}{\beta}\int_0^\beta d\lambda\textup{\textbf{Tr}}\rho_e  e^{\lambda H}X e^{-\lambda H} Y
\end{equation}
which is a real quantity  \cite{kubo1957statistical,kubo1959some}.
It is fundamental because the linear transport coefficients are computed integrating it in time, and its symmetries are reflected in the symmetries of the Onsager transport matrix. With this definition, the response function \eqref{2.60} takes the form 
\begin{equation}
\label{2.74}
   \phi_{BA}(t)= \textbf{Tr}\rho_e(A(0), B(t))=\langle(A(0), B(t))\rangle=\beta\langle \dot{A}(0);B(t)\rangle
\end{equation}
where we used \eqref{2.71} in the integral in \eqref{2.69}. 

Suppose the external force be either periodic or expressed in terms of Fourier modes:
\begin{equation}
\label{2.63}
    F(t)=Re(F_0 e^{i\omega t})
\end{equation}
Then, the Linear Response at each frequency $\omega$ is given by
\begin{equation}
\label{2.64}
    \mathbb{E}_t[\Delta B]=Re(\rchi_{BA}F_0e^{i\omega t}) \, ,
    ~~ \mbox{with } ~~ \rchi_{BA}(\omega)=\int_0^\infty \phi_{BA}(t)e^{-i\omega t}dt
\end{equation}
where $\rchi_{BA}$ is called \textit{admittance}, and using the above definitions can also be written as:
\begin{equation}
    \rchi_{BA}(\omega)=\beta\int_0^\infty\langle \dot{A}(0);B(t)\rangle e^{-i\omega t}dt \, .
\end{equation}

This theory is very general and accurate, as long as the perturbation is not excessively large, something not common on the thermodynamic scale. Given the perturbation and the observable of interest, everything reduces to the integration of the corresponding  correlator \cite{j2007statistical}. For instance, one obtains the linear transport coefficients as so-called {\em Green-Kubo} integrals, three examples of which are the following:
\begin{eqnarray}
\label{Dietc}
&&D = {1 \over 3} \int_0^\infty {\rm d} t \, \left\langle v(0) v(t) \right\rangle \, , \quad \mbox{for 3-dimensional self diffusion} \\
&&\lambda = {V \over 3 k_B T^2} \int_0^\infty {\rm d} t \, \mathbb{E}\left\langle J_q(0) J_q(t) \right\rangle \, , 
\quad \mbox{for thermal conductivity} \\
&&\eta = {V \over 3 k_B T} \int_0^\infty {\rm d} t \, \mathbb{E}\left\langle P_{xy}(0) P_{xy}(t) \right\rangle \, , 
\quad \mbox{for 2-dimensional shear viscosity}
\end{eqnarray}
where $V$ is the volume occupied by the system, and $T$ is its temperature, $v$
is the velocity of a test particle, $J_q$ the heat flux, and $P_{xy}$ the off-diagonal entry of the pressure tensor.
Both in classical and in quantum mechanics, the observables are in general expressed in function of coordinates and momenta, apart from the case of particle with spin that we will discuss later. 

Linear response theory, and the  Fluctuation-Dissipation theorem in particular, represent a cornerstone of Statistical Mechanics, expressing macroscopic \textit{nonequilibrium} properties in terms of {\em equilibrium} microscopic dynamics and time correlation functions. This provides an atomistic  justification, and a molecular dynamics computational tool for the very widely applicable linear relations such as Ohm, Fick and Fourier laws, the Navier-Stokes equations, etc.

\subsection{Onsager Reciprocal Relations}
In Refs.\cite{onsager1931reciprocal,onsager1931reciprocal2}, Onsager derived his celebrated Reciprocal Relations, for particles systems whose dynamics are time reversal invariant (TRI). These are symmetry relations for the linear transport coefficients that, in accord with the theory outlined above, are implied by symmetries of the corresponding correlation functions \eqref{2.74}, such as:
\begin{equation}
    \phi_{\varphi \psi}(t)= \eta_\varphi \eta_\psi \phi_{\psi \varphi}(t)
    \label{fiABsymm}
\end{equation}
where $\eta_\varphi , \eta_\psi = \pm 1$ are the parities of the functions $\varphi$ and $\psi$ under time reversal.
Classically, the canonical TRI corresponds to the invariance of the equations of motion under the action of the following map:
\begin{equation}
\label{39}
    \mathcal{M}_c(\boldsymbol{X},\boldsymbol{P})=(\boldsymbol{X},-\boldsymbol{P})
\end{equation}
where $\Gamma=(\boldsymbol{X},\boldsymbol{P})$ represents the coordinates and the conjugate momenta of a particle system, {\em i.e}\ one point in the corresponding phase space. Quantum mechanically, this property is expressed by the invariance of the Schr\"odinger equation under Wigner's time reversal operator $\mathcal{T}=UK$, where $K$ is the complex conjugation and $U$ is a unitary operator such that \cite{sachs}:
\begin{equation}
\label{40}
    \mathcal{T}^{-1}(\boldsymbol{X},\boldsymbol{P})\mathcal{T}=(\boldsymbol{X},-\boldsymbol{P})
\end{equation}
with $\boldsymbol{X}$ and $\boldsymbol{P}$
the position and the momentum operators.
Letting $S^t$ be the evolution operator,
which classically is the Hamiltonian flow in the phase space, and quantum mechanically is given by the unitary operator $S^t=e^{-iHt/\hbar}$ (for time independent Hamiltonians $H$), TRI amounts to the validity of:
\begin{equation}
    \mathcal{M}_cS^t=S^{-t}\mathcal{M}_c  ~~ \mbox{classical mechanics; } ~~~ {\cal T} S^t = S^{-t} {\cal T} ~~
    \mbox{quantum mechanics}
\end{equation}
In Onsager's words, TRI means that when
{\em "the velocities of all the particles present are reversed simultaneously the particles will retrace their former paths, reversing the entire succession of configurations"}. As mentioned above, this property was considered essential in the derivation of the Onsager Reciprocal \cite{onsager1931reciprocal},
but absent in presence of magnetic fields or rotating frames.
Hence, the Reciprocal Relations were  thought to be inapplicable.
Casimir then realized that a modification of the Onsager relations holds also in these cases if the magnetic field {\bf B} is inverted by the time reversal 
map \cite{Casimir}.\footnote{Because the case of rotating frames can be treated in the same way, from now on we only write about the magnetic field.} The result is that 
%
the symmetry
property \eqref{fiABsymm} is replaced by
the following:
\begin{equation}
    \label{eq:1}
    \phi_{\varphi\psi}(t,\mathbf{B}) = 
    \langle\varphi(0);\psi(t)\rangle_\mathbf{B}=\eta_\varphi\eta_\psi\langle\varphi(0);\psi(-t)\rangle_{-\mathbf{B}} =\eta_\varphi\eta_\psi\langle\varphi(t);\psi(0)\rangle_{-\mathbf{B}} = \phi_{\psi\varphi}(t,-\mathbf{B}) 
\end{equation}
where the angular brackets represent averaging with respect to the equilibrium with magnetic field $\mathbf{B}$ or $-\mathbf{B}$, as indicated in the formula.
This is substantially different from the original case, that concerned a single system, instead of two systems in two different magnetic fields.
\cite{kubo1957statistical,kubo1959some,kubo1966fluctuation}. 
The importance of the Onsager Reciprocal Relations is discussed in many classic texts, {\em e.g.}\ in Refs.\cite{callen1948application,miller1960thermodynamics}. Also, Refs.\cite{benenti2017fundamental, benenti2020power} pointed out that an engine working at the Carnot efficiency while delivering finite power could be obtained if the Onsager could be violated
\cite{casa}.

\subsection{Beyond canonical time reversal}
In Ref.\cite{rondoe}, Bonella, Ciccotti and Rondoni noticed that a transformation used
in the theory of shearing fluids, called Kawasaki map \cite{j2007statistical}, preserves the equations of motion of systems of
classical charged particles subject to a constant magnetic field {\bf B}, with Hamiltonian:
\begin{equation}
\label{1.51}
    H=\sum_{i=1}^N\left[\frac{(\boldsymbol{p}_i-q_i\boldsymbol{A}(\boldsymbol x_i))^2}{2m_i}\right]+\sum_{i>j}v(|\boldsymbol x_i-\boldsymbol x_j|)
\end{equation}
where $\mathbf{A}$ is the vector potential of $\mathbf{B}$, and $v$ is a central interaction potential. The map, with {\bf B} replacing the shear rate, writes:
\begin{equation}
    \mathcal{M}_{SL}(x_i,y_i,z_i,p_i^x,p_i^y,p_i^z,t; \boldsymbol B)=(x_i,-y_i,z_i,-p_i^x,p_i^y,-p_i^z,-t; \boldsymbol B)\quad\quad \forall i\in[1,N]
\end{equation}
and implies the validity of the Onsager relations. Indeed, it is readily seen that ${\cal M}_{SL}$ preserves the equations of motion when time is reversed. Then, denoting by $S^t$ the time evolution operator, one obtains $\mathcal{M}_{SL} S^t=S^{-t}\mathcal{M}_{SL}$, 
or $S^{-t} = \mathcal{M}_{SL} S^t \mathcal{M}_{SL}$. That means that for every phase point $\Gamma$, there is another point $\widehat{\Gamma}=\mathcal{M}_{SL} S^t \mathcal{M}_{SL} \Gamma$
whose forward time trajectory, in some sense does the opposite of the forward time trajectory of $\Gamma$. The two trajectories do not coincide in 
configuration space, if the standard time reversal symmetry does not hold, but that turns irrelevant when considering correlations of currents, and averaging over all trajectories.

This idea has been successively extended in various papers on classical mechanics TRI, 
which showed that the set of time reversal operations that hold in a magnetic field is quite large \cite{bonella2017time}. This allows sharp predictions about transport coefficients \cite{bonella2017time,rondom}. Analogous conclusions have been reached in the case of non-relativistic quantum mechanics, where, however, the effect of spin had not been considered \cite{rondoquantu}. In fact, the canonical definition of time reversal on spin \cite{sachs,toda1991statistical} unavoidably leads to T-symmetry breaking already for Pauli Hamiltonian (spin 1/2). The existence of non-standard time reversal symmetries can then be used in other statistical studies. For instance,
it has led to the extension of the validity of Fluctuation Theorems \cite{evans1993probability} to particle systems in magnetic and electric fields \cite{rondoFT1,rondoFT2}.

\section{\large Results outline}
In this section we provide a summary of the results proven in Section \ref{TheorySect}, in particular: 
\begin{enumerate}
    \item the characterization of time reversal operations for Classical and
    non-relativistic Quantum Mechanics systems, including spin 1/2 particles,  in presence of an external magnetic field;
    \item the extension of compatibility conditions between time reversal operations and magnetic field;
    \item the extension to quantum mechanics and to non-constant magnetic fields of analytical results about correlators \cite{bonella2017time}, directly derived from the application of more than one time reversal symmetry.
\end{enumerate}   
The experimental relevance of the Onsager reciprocal relations in absence of magnetic effects, is reviewed in a wide fraction of literature, see {\em e.g.}\ 
Refs.\cite{miller1960thermodynamics,miller1973onsager,letey1969movement,monroe2006onsager,gaeta1994phonons, miyatani1981experimental,rowley1978dufour,matthews2014experimental,kimura2007room,sola2019experimental},
which span from diffusion of various kinds, to quantum thermoelectric transport, and even spin and charge simultaneous transport. 
In particular, Ref.\cite{jacquod2012onsager}
investigates theory and possible experiments about the reciprocal relations in spin-Hall and inverse spin-Hall effects, in spin-injection, in magnetoelectric spin currents, Seebeck, Peltier, spin Seebeck, and spin Peltier effects, in systems with and without coupling to superconductors. Or, on the other hand, Ref.\cite{marklund2007dynamics} specifically analyzes linear response in the context of Pauli equation and plasma dynamics.

Our results predict that the onset of an external magnetic field breaks neither the Onsager symmetry nor the validity of the Fluctuation Theorem a priori. Moreover, they predict that certain cross transport coefficients of {\em e.g.}\ superionic conductors are not merely small, hence hard to measure, they actually vanish. In fact, Refs.\cite{bonella2017time,rondoFT1,casa} have already verified numerically some of these predictions, using the previously found time reversal symmetries.

\subsection{Characterization of generalized time reversal operations}
First we identify the general structure of a time reversal operation that preserves the equations of motion. We begin with the classical mechanics a system of $N$ independent charged particles, each with 3 degrees of freedom, confined within a box of side $2L$, and coupled to an external magnetic field, with Hamiltonian
    \begin{equation}
    \label{H1}
    H=\sum_{i=1}^N\left[\frac{(\boldsymbol{p}_i-q_i\boldsymbol{A}(\boldsymbol x_i))^2}{2m_i}\right]
\end{equation}
Our Theorem \ref{theo:1.1.2}
shows that a class of extended time reversal operations $\mathcal{M}$ for such systems is represented by $6N\times 6N$ matrices of the form:
\begin{equation}
    \mathcal{M}=
    \begin{pmatrix}
    A&0\\
    0&-A\\
    \end{pmatrix}
\end{equation}
with $A\in O(3N)$, the orthogonal group, and $A^2=I$, the identity matrix, so that $\mathcal{M}(\boldsymbol X,\boldsymbol P)=(A\boldsymbol X,-A\boldsymbol P)$. 
These constitute an infinite set of transformations, that includes the standard time reversal operation $(\boldsymbol X,\boldsymbol P) \mapsto (\boldsymbol X,-\boldsymbol P)$, where $A=I$, and includes a number
\begin{equation}
    \Tilde{\Delta}_{even}(M)=\sum_{r_2=0}^{\lfloor M/2\rfloor}\frac{M!\, 2^{M-2r_2}}{(M-2r_2)!\,r_2!} \, , \quad \mbox{with } ~ M=3N ,
\end{equation}
of permutations of coordinates, cf.\ Corollary \ref{cor:1.1.1} below.

For the non-relativistic spinless quantum systems, time reversal concerns the Schr\"odinger equation, and is represented 
by the Wigner operator $\mathcal{T}$ defined above, keeping in mind that a transformation of the coordinate operator $\boldsymbol X$ implies a modification of the momentum operator $\boldsymbol P=-i\partial_{\boldsymbol X}$.
We state that $\mathcal{T}$ must preserve the canonical commutation relations, as in Classical Mechanics the Poisson brackets must be preserved. Then, the matrix representation of the action of $\mathcal{T}$ on $\boldsymbol X$ and $\boldsymbol P$ must be symplectic \cite{anderson1994canonical}. 
Considering that $K$ acts as the antisymplectic $6N\times 6N$ matrix $\operatorname{diag}(I,-I)$, where $I$ is the $3N\times 3N$ identity matrix, we conclude that $U$ must be antisymplectic. 
Our Theorem \ref{theo:1.2.1} characterizes $U$ similarly to the classical mechanical reasoning, apart from the fact that quantum mechanics also allows $\mathcal{T}^2=-I$, not only $\mathcal{T}^2=I$.
Moreover,  a simple example shows that $U$ has to be real valued, for the Hamiltonian $\eqref{H1}$.


Finally, we consider time reversal invariance for particles of spin 1/2. In  textbooks, such as Sachs \cite{sachs}, the action of the Wigner operator $\mathcal{T}=(U_{c}\otimes U_{spin})K$, where $U_{spin}$ acts on Pauli matrices $\boldsymbol \sigma_i=(\sigma^i_x,\sigma^i_y,\sigma^i_z)$, is defined considering spin analogous to angular momentum, which means:
\begin{equation}
\label{axio}
    \mathcal{T}\boldsymbol\sigma\mathcal{T}=-\boldsymbol\sigma
\end{equation}
This implies $U_{spin}=\bigotimes_{i=1}^N \sigma_y^i$.
In Proposition \ref{prop:2.6}, we show that thanks to the generalized time reversal operations also condition \eqref{axio} can be relaxed. In particular, it suffices that $\mathcal{T}$ preserves the canonical commutation relation for the Pauli matrices algebra $\mathfrak{su}(2)$. 


\subsection{Extension of compatibility conditions}
In this part of the paper we extend the compatibility conditions of Carbone and Rondoni \cite{carbone2020necessary}
between the time reversal operations known at the time and both vector potential and magnetic field. These are the necessary and sufficient conditions for the Hamiltonian $H$ to be invariant under time reversal.
The compatibility condition for the vector potential, that is defined up to a gauge transformation, was given by
\begin{equation}
    \label{2.8}
        \mathcal{M}_m\boldsymbol A(\mathcal{M}_m\boldsymbol x)=-[\boldsymbol A(\boldsymbol x)]_R
    \end{equation}
where $\mathcal{M}_m$ is the representation of the time reversal map on the space of the coordinates of one particle, and $[\boldsymbol A(\boldsymbol x)]_R$ is the vector potential in the equivalence class of those with same magnetic field $\boldsymbol B=\nabla \times \boldsymbol A$. The compatibility condition with the magnetic field was given by
 \begin{equation}
        \label{2.9}
    \operatorname{det}(\mathcal{M}_m) \mathcal{M}_m\boldsymbol B(\mathcal{M}_m\boldsymbol x)=-\boldsymbol B(\boldsymbol x)
\end{equation} 
We extend these conditions obtained for classical systems to the generalized time reversal operations derived here, and to the quantum case, including the spin 1/2 particles.
In particular, 
given the Pauli Hamiltonian
\begin{equation}
   H=\sum_{i=1}^N{\frac {1}{2m_i}}\left[(\boldsymbol {p}_i -q_i\boldsymbol {A}(\boldsymbol x_i) )^{2}-q_i\boldsymbol{\sigma}_i\cdot\boldsymbol{B}(\boldsymbol x_i)\right] 
\end{equation}
we prove that, contrary to expectations that
the additional term of spin field coupling $\boldsymbol\sigma_i\cdot \boldsymbol B$ reduces the number of available time reversal operators, 
no additional compatibility condition other than \eqref{2.9} is needed.
In other words, 
if a time reversal operator preserves the spinless minimal coupling term, there is
an appropriate $U_{spin}$ that 
preserves also the spin field coupling. 
This result is fundamental because it means that from the point of view of statistical relations, such as the Onsager relations and the Fluctuation Theorem there is no effective difference between spinless and spin $1/2$ particles described via Pauli Hamiltonian.

Finally, we extend a result proven in \cite{bonella2017time} showing that 
the off diagonal entries of the diffusion tensor for classical particles subjected to a particular constant magnetic field obey:
\begin{equation}
\label{dd}
    D_{xy}=-D_{yx}
\end{equation}
We show that the same holds 
in quantum systems, and also in presence of a non-constant magnetic field along $z$ axis.

\section{\large Theory}
\label{TheorySect}
As recalled in Sect.\ref{sec:Introduction}, sufficient conditions for the validity of Onsager reciprocal relations are known, and they include a set of generalized time reversal symmetries. On the contrary, very little is known about \textit{necessary} conditions.
First, in Sec.\ref{ClassMe}, we recall the main facts about time reversal 
transformation in classical mechanics, introducing a whole new class of operations. Then, we illustrate quantum mechanical time reversal
operators, in the case of spinless particles, extending the treatment of Sec.\ref{ClassMe} to the quantum realm, cf.\ Sec.\ref{sec:The time reversal operator}. In Sec.\ref{sec:Time reversal with spin} we investigate the issue of time reversal in the context of quantum systems composed by particle of spin $1/2$. In Sec.\ref{sec:compatib} we summarize the concept of Kubo canonical quantum correlators, as well as the derivation of Onsager relations from a generalized time reversal symmetry obtained in Ref.\cite{rondoquantu}. 
Subsequently, we illustrate the compatibility conditions between the different time reversal operations and a generic magnetic field. We recall the sufficient conditions of Ref.\cite{carbone2020necessary} that lead to Onsager relations in classical mechanics, and we extend them to the operations found in Sec.\ref{ClassMe} and to the case of particles with spin. The striking result of this Section is that a time reversal operator that works in the spinless case, works also for particles of spin $1/2$. Finally, in Sec.\ref{sec:example} we study the constraints imposed by TRI on the diffusion tensor in a particular physical setup, in the wake of the reasoning of Ref.\cite{rondom}.

\subsection{Classical mechanics}
\label{ClassMe}
The microscopic state of a system made of $N$ classical particles is represented by the collection of coordinates and momenta of all its particles, that constitutes a point $\Gamma$ in the phase space $\Omega$. In case of dynamics determined by a Hamiltonian $H$, the time evolution of the coordinates and momenta is prescribed by the following equations of motion:
\begin{equation}
\label{eq:2.2}
\begin{dcases}
\frac{\partial H}{\partial p_i}=\dot{x}^i\\
\frac{\partial H}{\partial x^i}=-\dot{p_i}
\end{dcases}\;\;\;\;\;\; i=1,...,3N
\end{equation} 
Denote by $S^t : \Omega \to \Omega$ the operator that expresses the solutions of the equations of motion up to time $t$, {\em i.e.}\ $S^t \Gamma$ is the state at time $t$, if it was $\Gamma$ a time 0. Following Abraham Ref.\cite{abraham1978foundations}, the dynamics are called time reversal invariant (TRI) if there exists a linear antisymplectic involution 
${\cal M} : \Omega \to \Omega $ (an operator such that ${\cal M}^2 = I$, the identity) such that:
\begin{equation}
    {\cal M} S^t \Gamma = S^{-t} {\cal M} \Gamma \, ,
    \quad \forall \Gamma \in \Omega \, , ~ \forall t \in \mathbb{R}
    \label{TRIdef}
\end{equation}
The reason for this terminology is that right multiplication by $\cal M$ before application to $\Gamma$ yields:
\begin{equation}
    S^{-t} = {\cal M} S^t {\cal M}
    \label{TRIjusti}
\end{equation}
which means that the backward time evolution is conjugated to the forward time evolution {\em via} the involution $\cal M$, in such a way that one backward trajectory can be obtained by properly applying $\cal M$ to the forward trajectory.

Microreversibility corresponds to TRI under the involution $\mathcal{M}$ defined by ${\cal M}({\bf x},{\bf p})=({\bf x},-{\bf p})$,
but we call time reversal invariant
all the dynamics for which one involution obeying Eq.\eqref{TRIdef} exists.

From the point of view of the equations of motion, we can define an extended operation  $\mathcal{T}\equiv\mathcal{M}\circ\mathbb T$, where $\mathbb{T}$ maps the time parameter $t$ to $-t$ in Eqs.\eqref{eq:2.2}. The system verifies TRI under $\mathcal{M}$ if the application of $\mathcal{T}$ to the equations of motion leaves them unchanged. Note that
the concept of time reversal defined by Eq.\eqref{TRIdef} concerns trajectories in phase space.
\begin{defi}
\label{sym}
A matrix operator $P$ is symplectic or, respectively,
antisymplectic on a $2n$-dimensional phase space $\Omega$ if 
\begin{equation}
    P^T\omega P=\omega \quad \mbox{\rm or } ~~
    P^T\omega P=-\omega
\end{equation}
where $\omega$ is the antisymmetric matrix
\begin{equation}
\label{anti}
    \omega=
    \begin{pmatrix}
        0&-I_{n}\\
        I_n&0
    \end{pmatrix}
\end{equation}
and $I_n$ is the $n\times n$ identity matrix. 
\end{defi}
\noindent
This can now be used to define the time reversal transformations of Hamiltonian systems.
\begin{defi}
\label{defi:1.1.2}
A time reversal transformation $\mathcal{M}$, for a Hamiltonian system, is a linear operator acting on the space $\Omega$ in such a way that:
\begin{itemize}
    \item It is an involution, that is $\mathcal{M}^2=I$.
    \item It is an antisymplectic linear operator.
\end{itemize}
\end{defi}
As in Ref.\cite{carbone2020necessary}, we begin considering transformations that separately act on the $6$-dimensional subspaces of $\Omega$ that concern single particles. In this case, a time reversal transformation on
$\Omega$ can be written as:
\begin{equation}
\label{eq:2.3}
\mathcal{M}(x,y,z,p_x,p_y,p_z)=P(s_1x,s_2y,s_3z,-s_1p_x,-s_2p_y,-s_3p_z)
\end{equation}
where $s_i=\pm1$, and $P$ is a permutation obeying $P^2=I$, which acts in the same way on coordinates and on the corresponding momenta. Our first result is the following theorem.
\begin{theo}
\label{theo:1.1.2}
Consider a system of $N$ particles subject to an external magnetic field, described by the Hamiltonian
\begin{equation}
\label{hamii}
    H=\sum_{i=1}^N\left[\frac{(\boldsymbol{p}_i-q_i\boldsymbol{A}(\boldsymbol x_i))^2}{2m_i}\right]
\end{equation}
The matrix representation of the antisymplectic operator $\mathcal{M}$ representing a time reversal operation for such a system is block diagonal on the phase space $\Omega$ and takes the form:
\begin{equation}
    \label{1.1.30}
    \mathcal{M}=
    \begin{pmatrix}
    A&0\\
    0&-A\\
    \end{pmatrix}
\end{equation}
    with $M=3N$, $A\in O(M)$ and $A^2=I$.
\end{theo}
\begin{proof}
We start from a generic matrix $\mathcal{M}\in GL(2M,\mathbb{R})$, {\em i.e.}
\begin{equation}
\label{1.1.31}
    \mathcal{M}=
    \begin{pmatrix}
    A&B\\
    C&D
    \end{pmatrix}
\end{equation}
where $A$, $B$, $C$ and $D$ are $M\times M$ blocks. As stressed in Refs.\cite{rondoe,carbone2020necessary}, a time reversal operation defined as in Definition \ref{defi:1.1.2} yields TRI if and only if
\begin{equation}
    H(\mathcal{M}\Gamma)=H(\Gamma) \, , \quad \forall \, \Gamma \in \Omega
\end{equation}
Thus, the Hamiltonian \eqref{hamii} requires $\mathcal{M}$ to not mix coordinates and momenta. In the case it does, the minimal coupling term $(\boldsymbol{p}_i-q_i\boldsymbol{A}(\boldsymbol x_i))^2$ is not preserved by the operation. In formulae, the matrix must be block diagonal:
\begin{equation}
\label{1.1.32bis}
\mathcal{M}=
    \begin{pmatrix}
    A&0\\
    0&D
    \end{pmatrix}
\end{equation}
As $\mathcal {M}$ must preserve the norm in the $2N$-dimensional phase space, because of the presence of the square in the minimal coupling term, $\mathcal{M}$ must be orthogonal and, consequently one has $A\in O(M)$ and $D\in O(M)$, with $M=3N$.
Moreover, the transformation must be antisymplectic by Definition \ref{defi:1.1.2}. Then, we impose the constraint:
\begin{equation}
\label{1.1.33}
\mathcal{M}^T\omega\mathcal{M}=-\omega \, ; \quad \mbox{i.e.}\ ~~
\begin{pmatrix}
    A^T&0\\
    0&D^T
    \end{pmatrix}
     \begin{pmatrix}
        0&-I_n\\
        I_n&0
        \end{pmatrix}
    \begin{pmatrix}
    A&0\\
    0&D
    \end{pmatrix}
    =
    \begin{pmatrix}
        0&I_n\\
        -I_n&0
        \end{pmatrix}
\end{equation}
which leads to the constraint $A^TD=-I$. Then,
\begin{equation}
\label{1.1.34bis}
\mathcal{M}=
    \begin{pmatrix}
    A&0\\
    0&-(A^T)^{-1}
    \end{pmatrix}
\end{equation}
Finally, because $\mathcal{M}$ is block diagonal, and $\mathcal{M}^2=I$,  Eq.\eqref{1.1.32bis} implies $A^2=I$ and $D^2=I$. Ultimately, since $A$ is orthogonal, that is $AA^T=I$, and $A^2=I$, we obtain that $A$ is a symmetric matrix, {\em i.e.} $A=A^T$.
The proof is then complete, observing that
\begin{equation}
    \label{1.1.35}
    (A^T)^{-1}=A^{-1}=A
\end{equation}
\end{proof}
This means that the number of time reversal operations is the number of matrices $A$ that are orthogonal and involutory. In particular, let us consider $3N \times 3N$ matrices whose entries are $0$ or $1$, called binary matrices (more general matrices will be considered later, in Section \ref{sec:compatib}). Among binary matrices we have the following:
\begin{defi}
A permutation matrix is a matrix obtained permuting the row of an identity matrix of the same dimension.
\end{defi}

\noindent
We are now going to find the time reversal matrices using some results of Group Theory \cite{tung1985group}. First, note the following proposition.
\begin{prop}
\label{prop:2.7}
The involutory binary matrices $A$ are in one-to-one correspondence with the linear representations of the cyclic group $\mathbb{Z}_2$ over a $3N$-dimensional vector space.
\end{prop}
\begin{proof}
By definition, the group $\mathbb{Z}_2$ is the set $\{e,a\}$ with $e$ the group identity and $a^2=e$. A linear representation $\mathcal{R}(a)$ on a $3N$-dimensional vector space is a $3N\times3N$ matrix such that $\mathcal{R}(a)^2=\mathcal{R}(e)=I_{3N}$. 
Moreover, $S_2$, the symmetric group on $2$ elements, coincides with $\mathbb{Z}_2$ and, in general, a representation of an element of $S_k$ must be a permutation, hence an orthogonal, matrix. Then, a matrix $\mathcal{R}(a)$ has the required property, and the possible $A$ and the $\mathcal{R}(a)$ are in one-to-one correspondence.
\end{proof}

Considering that one such matrix $A$ is a permutation, it is a representation of an element of $S_{3N}$, like $\mathcal{R}(a)$. Now, let us introduce the cycle decomposition of a permutation of $S_{3N}$, denoted by:
\begin{equation}
\label{cycle}
\underbrace{(\cdot)...(\cdot)}_{r_1}\underbrace{(\cdot\cdot)...(\cdot\cdot)}_{r_2}...\underbrace{(\cdot...\cdot)}_{r_{3N}}
\end{equation}
which is invariant under group conjugation, and allows us to uniquely label each conjugation class with a set of integers $\{r_1,...,r_{3N}\}$ satisfying the following constraint:
\begin{equation}
\label{1.1.49}
  \sum_{l=1}^{3N} lr_l=3N 
\end{equation}
There is a useful way to represent the conjugation classes:
\begin{defi}[Young tableaux]
\label{defi:1.1.6}
There is a one-to-one correspondence between a conjugation class $\{r_l\}$ of the group $S_k$ and the following graphical representations known as Young tableau,
\begin{equation}
   \label{1.1.52}
\ytableausetup
{mathmode, boxsize=2em}
\begin{ytableau}
\scriptstyle 1 & \scriptstyle 2 & \scriptstyle 3 & \none[\dots]
& \scriptstyle a_1- 1 & \scriptstyle a_1 \\
\scriptstyle 1 & \scriptstyle 2& \none[\dots]& \none[\dots]
& \scriptstyle a_{2} \\
\none[\vdots] & \none[\vdots]
& \none[\vdots] \\
\scriptstyle 1 & \scriptstyle a_{k-1} \\
\scriptstyle a_k
\end{ytableau}
\end{equation}
where the number of squares $a_l$ in each of the $k$ rows obeys the following rules:
\begin{enumerate}
    \item $a_{l+1}\leq a_{l}\;\;\;\forall l=1,...,k$.
    \item The total number of squares in the tableau equals $k$: $\sum_{l=1}^k a_l=k$
\end{enumerate}
Then, the one-to-one correspondence between the class $\{r_l\}$ and the corresponding Young tableau is given by:
\begin{equation}
       \label{1.1.53}
       \begin{cases}
       r_l=a_l-a_{l+1}\; , \;\;\;\forall \, l=1,...,k-1\\
       r_k=a_k
       \end{cases}
\end{equation}
\end{defi}
\noindent
For instance, $S_3$ can be associated with any of the three following Young tableaux:
\begin{equation}
    \yng(3) \qquad \qquad \;\;\;\;\yng(2,1)\;\;\;\; \qquad \qquad \yng(1,1,1)
\end{equation}
that respectively correspond to three sets $\{r_l\}$: $\{3,0,0\}$, $\{1,1,0\}$ and $\{0,0,1\}$. Incidentally, this also constitutes one method for identifying the partitions of a fixed integer $n$ (see Cap. 5.2 in Ref.\cite{tung1985group}).
Now, the constraint \eqref{1.1.49} takes an interesting form, thanks to the following Corollary.
\begin{cor}
\label{cor:2.1}
The conjugation classes $\{r_i\}$ of $S_{3N}$ containing elements represented by involutory binary matrices obey
\begin{equation}
     \label{constra}
     r_l=0\;\;\;\;\; \forall l\geq 3
\end{equation}
\end{cor}
\begin{proof}
This is a consequence of Proposition \ref{prop:2.7}: the conjugation class represented by $A$ contains elements $a$ such that $a^2=e$, that is its cycle decomposition can be made only of cycles of order $1$ or $2$.
\end{proof}
\noindent
To compute the number $|\{r_i\}|$ of elements in a conjugation class $\{r_i\}$, we use the following known fact \cite{tung1985group}:
\begin{theo}
The number of elements in the conjugation class $\{r_l\}$ is
\begin{equation}
    \label{1.1.51}
    |\{r_l\}|=\frac{k!}{\prod_{i=1}^k i^{r_i}r_i!}
\end{equation} 
\end{theo}
\noindent
which leads us the fundamental result of this Section:
\begin{cor}[Number of generalized time reversal operations]
\label{cor:1.1.1}
Setting $M=3N$,
the maximum number of generalized time reversal operations involving binary matrices, for a system of $N$ particles described by the Hamiltonian \eqref{hamii}, is given by: 
\begin{equation}
     \label{1.1.56}
    \Tilde{\Delta}_{even}(M)=\sum_{r_2=0}^{M/2}\frac{M!\, 2^{M-2r_2}}{(M-2r_2)!\,r_2!}
\end{equation}
if $M$ is even and by
\begin{equation}
     \label{1.1.57}
    \Tilde{\Delta}_{odd}(M)=\sum_{r_2=0}^{(M-1)/2}\frac{M!\, 2^{M-2r_2}}{(M-2r_2)!\,r_2!}
\end{equation}
if $M$ is odd.
\end{cor}
\begin{proof}
The total number of time reversal operations that satisfy Corollary \ref{cor:2.1} is the sum of the number of elements in the conjugation classes, {\em i.e.}
\begin{equation}
     \label{1.1.58}
     \tilde{\Delta}\equiv \sum_{\{r_l\}}|\{r_l\}|
\end{equation} 
under the condition \eqref{constra} which, together with \eqref{1.1.49}, yields:
\begin{equation}
     \label{1.1.59}
     r_1+2r_2=M
\end{equation} 
Then, using \eqref{1.1.51} and \eqref{1.1.59} in \eqref{1.1.58} we obtain 
\begin{equation}
    \label{1.1.60}
    \Delta_{even}(M)=\sum_{r_2=0}^{M/2}\frac{M!}{(M-2r_2)!r_2!\,2^{r_2}}
\end{equation}
for $M$ even, while
\begin{equation}
    \label{1.1.61}
    \Delta_{odd}(M)=\sum_{r_2=0}^{(M-1)/2}\frac{M!}{(M-2r_2)!r_2!\,2^{r_2}}
\end{equation}
for $M$ odd, where the sum stops at $(M-1)/2$, because of the constraint in Eq.\eqref{1.1.59}. 

Now, consider for instance one conjugation class $\{r_l\}$, hence a single addend of the summation in these formulae. Each cycle in a decomposition like \eqref{cycle} is defined up to a minus sign. For example, take the matrix representation 
\begin{equation}
    \begin{pmatrix}
        0&s_p&0\\
        s_P&0&0\\
        0&0&s_3
    \end{pmatrix}
\end{equation}
associated to the time reversal operation that acts as a block diagonal on each particle subspace as
\begin{equation}
\mathcal{M}(x,y,z,p_x,p_y,p_z)=(s_Py,s_Px,s_3z,-s_Pp_y,-s_Pp_x,-s_3p_z)
\end{equation}
That corresponds to a permutation belonging to the conjugation class $\{r_l\}=\{1,1,0\}$ of $S_3$: it is evident as the $2\times2$ block that corresponds to the cycle of order $2$ carries a sign $s_P$, and similarly the cycle of order $1$ for $s_3$. In general, we can multiply by $\pm 1$ each cycle block in the representative matrix. This proves that we have to multiply each addend of the summations in \eqref{1.1.60} and \eqref{1.1.61} by a factor $2^{r_1+r_2}$, obtaining:
\begin{equation}
    \label{1.1.62}
    \tilde{\Delta}_{even}(M)=\sum_{r_2=0}^{M/2}\frac{M!\,2^{r_1+r_2}}{(M-2r_2)!r_2!\,2^{r_2}}
\end{equation}
\begin{equation}
    \label{1.1.63}
    \tilde{\Delta}_{odd}(M)=\sum_{r_2=0}^{(M-1)/2}\frac{M!\,2^{r_1+r_2}}{(M-2r_2)!r_2!\,2^{r_2}}
\end{equation}
Finally, recalling relation \eqref{1.1.59} we get the two formulae \eqref{1.1.56} and \eqref{1.1.57}.
\end{proof}

\noindent
To illustrate this result, let us apply it to the $6$-dimensional subspace of a system with $N=1$. Here, $M=3$ so we use \eqref{1.1.57} and we obtain:
\begin{equation}
    \label{1.1.64}
    \Tilde{\Delta}_{odd}(3)=\sum_{r_2=0}^{1}\frac{M!\, 2^{M-2r_2}}{(M-2r_2)!\,r_2!}=\frac{3!\, 2^{3}}{3!\,0!}+\frac{3!\, 2^{1}}{1!\,1!}=8+12=20
\end{equation}
Correctly, this result coincides with the one obtained in Ref.\cite{carbone2020necessary} for single particle subspaces of $N$ particle systems. This kind of operations swap different coordinates and momenta of a given particle. For example, they include the time reversal transformation
\begin{equation}
    \label{nondiag}
    ({x}_1,...,{x}_j,{x}_{j+1},...,{x}_M,{p}_1,....,{p}_j,{p}_{j+1},...,{p}_M)\xrightarrow{\mathcal{M}_{nd}}({x}_1,...,{x}_{j+1},{x}_j,...,{x}_M,-{p}_1,....,-{p}_{j+1},-{p}_j,...,-{p}_M)
\end{equation}
that permutes coordinates $x_j$ and $x_{j+1}$ of a given particle. In fact, this operation corresponds to a matrix $A$ of form
\begin{equation}
\begin{pmatrix}
    I_{j-1}&0&0\\
    0&B&0\\
    0&0&I_{M-j-1}
\end{pmatrix}
\end{equation}
where $I_k$ is the $k\times k$ identity matrix and $B$ is a $2\times 2$ block such as
\begin{equation}
B=
\begin{pmatrix}
    0&1\\
    1&0
\end{pmatrix}
\end{equation}
This matrix is the representation of a permutation that belongs to the conjugation class $\{M-2,1\}$, that is the permutations composed by $M-2$ cycles of order $1$ and one cycle of order $2$.

As last remark, the generalization of the concept of time reversal leads to a notable outcome on the rule of transformation of other classical observables, as for example angular momentum. Let us recall its definition:
\begin{defi}
For a classical Hamiltonian system, the total angular momentum is defined by 
\begin{equation}
\label{2.62}
    \boldsymbol L=\sum_{i=1}^{N}\boldsymbol x^i\times\boldsymbol p_i
\end{equation}
\end{defi}
\noindent
The canonical time reversal operation $(\boldsymbol X,\boldsymbol P) \mapsto (\boldsymbol X,-\boldsymbol P)$ trivially transforms $\boldsymbol L$ into $-\boldsymbol L$. The same holds for the $20$ time reversal operations that do not permute coordinates of different particles. Restricting to the single particle subspaces $(\boldsymbol x^i,\boldsymbol p_i)$, and denoting the transformation rule by:
\begin{equation}
    (\boldsymbol x^i,\boldsymbol p_i)\to(R\boldsymbol x^i,-R\boldsymbol p_i)
\end{equation}
where Theorem \ref{theo:1.1.2} implies $R\in O(3)$, we obtain:
\begin{equation}
    \boldsymbol L \mapsto \boldsymbol L'=-\sum_{i=1}^{N}R\boldsymbol x^i\times R\boldsymbol p_i=-\boldsymbol L
\end{equation}
where we used the fact that the cross product is unchanged by coherent rotation of both factors. 
On the other hand, transformations with matrix representation like \eqref{1.1.30}, where $A$ does not separately act on each particle subspace, do not reverse $\boldsymbol L$. Consider, for example, the following generalized time reversal operation:
    \begin{equation}
\label{1.1.39}
     ({x}_1,...,{x}_j,{x}_{j+1},...,{x}_{3N},{p}_1,...,{p}_j,{p}_{j+1},...,{p}_{3N})\xrightarrow{\mathcal{M}_{nd}}({x}_{3N},...,{x}_{j},{x}_{j+1},...,{x}_1,-{p}_{3N},...,-{p}_{j},-{p}_{j+1},...,-{p}_1)
\end{equation}
which swaps $x_1$ and $x_{3N}$ and coherently acts on momenta as prescribed by Theorem \ref{theo:1.1.2}. Then, consider the $z$ component of $\boldsymbol L$. By definition of cross product 
we have:
\begin{equation}
    L_z=x^1p_2-x^2p_1+...+x^{3N-2}p_{3N-1}-x^{3N-1}p_{3N-2}
\end{equation}
which becomes
\begin{equation}
    L_z=-x^{3N}p_2+x^2p_{3N}+...-x^{3N-2}p_{3N-1}+x^{3N-1}p_{3N-2}\neq -L_z
\end{equation}
under the application of $\mathcal{M}_{nd}$. The same holds for $L_x$ and $L_y$. The conclusion is the following:
\begin{prop}
\label{prop:2.6}
There exist generalized time reversal operations that do not reverse the sign of the classical mechanics angular momentum.
\end{prop}
\noindent
Note that the time reversal operation $\mathcal{M}_{nd}$ works, in particular for a system of free equal mass particles, whose Hamiltonian is expressed by $H=\sum_{i=1}^{3N} p_i^2/2m$.

We now extend to Quantum Mechanics, the present reasoning. Indeed, in Ref.\cite{rondoquantu}, only $8$ time reversal operations had been identified, that classically take the form 
\begin{equation}
\mathcal{T}(x,y,z,p_x,p_y,p_z,t)=(s_1x,s_2y,s_3z,-s_1p_x,-s_2p_y,-s_3p_z,-t)
\end{equation}

\subsection{The time reversal operator}
\label{sec:The time reversal operator}
Let us begin with spinless non-relativistic particles. The fundamental axiom is that the state of the system is represented by a wave function, that obeys the Schr\"odinger equation: 
\begin{equation}
    \label{eq:2.1}
    i\frac{\partial\psi(x,y,z,t)}{\partial t}=H\psi(x,y,z,t)
\end{equation}
where our units imply $\hslash=1$. 
As in Classical Mechanics, we have an evolution equation that is invariant under time reversal. The corresponding Hilbert space is the quantum counterpart of the classical phase space. Then, we follow  Wigner's approach to define time reversal transformations, cf.\ Refs.\cite{wigner1932operation,sachs}. In the following we will consider operators and their matrix representations; for the sake of simplicity, 
the same symbol will be used.
    \begin{defi}[Spinless particles]
    \label{prop:1.2.1}
    An operator on the Hilbert space of the wave functions of an $N$ particle system is called a time reversal operator $\mathcal{T}$ if it obeys:
    \begin{itemize}
    \item $\mathcal{T}=UK$, with $U$ a unitary operator and $K$ the complex conjugation operator, \textit{i.e.} it is antilinear
    \item It is an involution in the quantum sense, that is $\mathcal{T}^2=\pm I$, where $I$ is the identity operator
    \item It is kinematically admissible, that is it preserve the canonical commutation relations $[x_n,p_m]=i\delta_{nm}$
    \end{itemize}
    \end{defi}
\noindent
To proceed analogously to the classical case, we exploit the following result \cite{anderson1994canonical}:
\begin{prop}
\label{prop:1.2.3}
The matrix representation of a time reversal operator $\mathcal{T}$ 
for a system of $N$ particles belongs to 
$Sp(6N, \mathbb{C})$.
\end{prop}
\noindent
For instance, in the case of a system with only two coordinates, the operators $x_1$ and $x_2$, and of a time reversal operator such that:
\begin{equation}
    \begin{split}
        \mathcal{T}x_1\mathcal{T}^{-1}=x_2\\
        \mathcal{T}x_2\mathcal{T}^{-1}=x_1\\
    \end{split}
\end{equation}
the matrix representation of $\mathcal{T}$ in terms of coordinates is given by:
\begin{equation}
    \begin{pmatrix}
        0&1\\
        1&0
    \end{pmatrix}
\end{equation} 
The Stone-Von Neumann Theorem \cite{von1931uniqueness} now states that in coordinates representation, {\em i.e.} for a wave function defined as coordinate dependent, the operators $x^i$ and $p_i$  uniquely act as:
\begin{equation}
    \label{1.89}
    \begin{dcases}
        x^i\psi(x_0)=x_0^i\psi(x_0)\\
        p_i\psi(x_0)=-i\frac{\partial \psi}{\partial x^i}(x_0)
    \end{dcases}
\end{equation}

\noindent
One can equivalently define the action of coordinate and momentum operators in  momentum space, via Fourier transform.

\begin{prop}
\label{prop:2.3b}
The complex conjugation preserves the canonical commutation relation.
\end{prop}
\begin{proof}
Using Eq. \eqref{1.89}, we can obtain the rule of transformation of the coordinate and momentum operators:
\begin{equation}
\label{2.42}
    K p_i K^{-1}=-KiK^{-1}\dfrac{\partial }{\partial x_i}=i \dfrac{\partial }{\partial x^i}=-p_i
\end{equation}
and 
\begin{equation}
\label{2.43}
    Kx^iK=x^i
\end{equation}
The proof of the canonical nature of $K$ is then trivial, since applying $K$ to both side of the canonical commutation relations
\begin{equation}
    \label{1.100c}
    \left[x^i,p_j\right]=i \delta^i_j
\end{equation}
one obtains
\begin{equation}
    \label{1.100b}
    \left[Kx^iK,Kp_jK\right]=KiK \delta^i_j
\end{equation}
that is equivalent to \eqref{1.100c} by definition of $K$.
Then, the $6N\times6N$ matrix representing $K$ on the $6N$-dimensional space of symbolic vectors $(\boldsymbol X,\boldsymbol P)$, is given by:
\begin{equation}
\label{krepre}
K=
    \begin{pmatrix}
    I&0\\
    0&-I
    \end{pmatrix}
\end{equation}
which is trivially antisymplectic, and acts
separately on coordinate and momentum operators
\end{proof}
Counting the generalized time reversal operations is now reduced to finding alternatives to the matrix representation of $K$. Table \ref{tab:my_label} illustrates the parallel with Classical Mechanics.
\begin{table}[h]
    \centering
    \begin{tabular}{c|c|c}
       & Classical Mechanics &  Quantum Mechanics\\
       & & \\
       Time Reversal   & $\mathcal{M}$ &$\mathcal{T}=UK$\\
       & & \\
       Commutation Relations & $\{x^i,p_j\}=\delta^i_j$ &$\left[x^i,p_j\right]=i \delta^i_j$\\
    \end{tabular}
    \caption{Comparative scheme}
    \label{tab:my_label}
\end{table}

\noindent
Recall that a fundamental aspect of Classical mechanics is that $\mathbb{T}$ acts on the Hamiltonian structure as an antisymplectic operator, and then $\mathcal{M}$ is also antisymplectic in order to obtain a symplectic transformation. Now, Proposition \ref{prop:1.2.3} says that in Quantum Mechanics $\mathcal{T}$ is represented by a symplectic matrix. Furthermore, $K$ is represented by an antisymplectic matrix, that is it plays the role of $\mathbb{T}$.
So, the situation is very similar to 
the classical one, and $U$ is antisymplectic.
\begin{theo}
\label{theo:1.2.1}
Take the $3N$-dimensional vectors $\boldsymbol X$ and $\boldsymbol P$ of respectively the coordinate and momentum operators of a quantum system made of $N$ particles. Given a time reversal operator $\mathcal{T}=UK$, the matrix representation of $U$ on the $6N$-vectors $(\boldsymbol X,\boldsymbol P)$ takes the form
\begin{equation}
\label{1.102}
U=
    \begin{pmatrix}
    A&0\\
    0&-A
    \end{pmatrix}
\end{equation}
where $A$ is a symmetric or an antisymmetric matrix.
\end{theo}
\begin{proof}
As in Theorem \ref{theo:1.1.2}, $\mathcal{T}$ cannot swap coordinates and momenta of different particles, because of the minimal coupling term in the Hamiltonian. Then, the matrix representation of $U$ takes the form
\begin{equation}
U=
    \begin{pmatrix}
    A&0\\
    0&B
    \end{pmatrix}
\end{equation}
As in the proof of Theorem \ref{theo:1.1.2}, we obtain $A^\dagger B=-I$. Then, we can write:
\begin{equation}
U=
    \begin{pmatrix}
    A&0\\
    0&-(A^\dagger)^{-1}
    \end{pmatrix}
\end{equation}
where we used the fact that the antisymplectic condition for $Sp(6N,\mathbb{C})$ is expressed by:
\begin{equation}
    U^\dagger \omega U=-\omega
\end{equation} and involves the dagger operation instead of the transpose. Now, $U$ is unitary by Definition \ref{prop:1.2.1}, and so we have  $A=(A^\dagger)^{-1}$, which leads to Eq. \eqref{1.102}.

To prove that $A$ is symmetric or antisymmetric, use the involutory property $\mathcal{T}^{2}=UKUK=UU^*=\pm I$, which implies $AA^*=\pm I$. Moreover, $U$ is unitary, hence $AA^\dagger=I$. Then, for $AA^* = I$
we obtain $A^T=A$, {\em i.e.} $A$ symmetric, whilst for $AA^*=- I$ we have $A^T=-A$, and is $A$ antisymmetric.
\end{proof}
Note that $K$ is of the form \eqref{1.102}, and corresponds to the the canonical time reversal operation, that preserves coordinates and reverses momenta. 
If we restrict to the part of $U$ acting on the coordinate (and then separately on momentum) operators, we find that a complex transformation of $\boldsymbol X$ and $\boldsymbol P$ cannot be used. Suppose that 
\begin{equation}
    \mathcal{T}\boldsymbol P\mathcal{T}^{-1}=P'=\mathfrak{Re}(\boldsymbol P')+\mathfrak{Im}(\boldsymbol P')i
\end{equation}
and consider a system of free particles with  Hamiltonian $H=\boldsymbol P^2/2m$. Applying the time reversal operator, we have:
 \begin{equation}
    H'=\mathcal{T}H\mathcal{T}^{-1}=\frac{(\boldsymbol P')^2}{2m}=\frac{\mathfrak{Re}(\boldsymbol P')^2+2\mathfrak{Im}(\boldsymbol P')\mathfrak{Re}(\boldsymbol P')i-\mathfrak{Im}(\boldsymbol P')^2}{2m}
\end{equation}
where the Hamiltonian is preserved only if
$2\mathfrak{Im}(\boldsymbol P')\mathfrak{Re}(\boldsymbol P')=0$, since the $H$ is real in $\boldsymbol X$ and $\boldsymbol P$. 
But if $\mathfrak{Im}(\boldsymbol P')\ne0$, as
assumed, that requires $\mathfrak{Re}(\boldsymbol P')=0$. But then $H'<0$ should equal $H>0$, which is absurd. This reasoning can be extended to the case of an external magnetic field. Because $H\propto \sum_{i=0}^N (\boldsymbol p_i-q\boldsymbol A(\boldsymbol x_i))^2$, there is again an addend proportional to $\boldsymbol P^2$ and the proof applies. Therefore, we may adopt the following:
\begin{assu}
The unitary operation $U$ associated to a time reversal operator $\mathcal{T}=UK$ acting on the coordinate (and momentum) operators is real.
\end{assu}

\noindent
This assumption immediately leads to:
\begin{cor}
If $A$ is real and symmetric, it belongs to $O(3N)$, as in Classical Mechanics.
\end{cor}
\begin{proof}
In the real case, 
the constraints $AA^\dagger=I$ and $A=A^T$ imply $AA^T=I$, which means that $A\in O(3N)$.
\end{proof}

\noindent
We can then straightforwardly repeat the arguments developed for the classical case, starting from Proposition \ref{prop:2.7}.
In particular, we consider permutation matrices also in this quantum framework.
Therefore, the generalized time reversal operators can be counted using Corollary \ref{cor:1.1.1}. Moreover, let $A$ be real but antisymmetric. In this case, Theorem \ref{theo:1.2.1} requires $A^2=-I$. 
Then, consider the following definition.

\begin{defi}
Let $\mathbb{Z}_4$ be a finite cyclic group with generator $a$ ($a^4$ is the group identity $e$), and with the real $2\times 2$ matrix representation given by:
\begin{equation}
    \mathcal{R}_4(e)=I\quad\quad \mathcal{R}_4(a)=\begin{pmatrix}
    0&-1\\
    1&0
    \end{pmatrix}
    \quad\quad \mathcal{R}_4(a^2)=-I\quad\quad \mathcal{R}_4(a^3)=\begin{pmatrix}
    0&1\\
    -1&0
    \end{pmatrix}
\end{equation}
\end{defi}

\noindent
Then, the following holds.
\begin{prop}
\label{prop:1.2.6}
If the number of particles $N$ is odd, $A$ is not real and antisymmetric. If $N$ is even, the number of time reversal operators $\tilde{\Delta}$ (that is the number of possible $A$'s) equals
\begin{equation}
\label{1.108}
    \tilde{\Delta}=\frac{M!}{(M/2)!} \, \quad \mbox{with } ~~ M = 3N
\end{equation}
\end{prop}
\begin{proof}
The condition $A^2=-I$ means that $A\in GL(3N,\mathbb(R))$ is a linear representation of the element $a$ or $a^3$ of the group $\mathbb{Z}_4$ over a $3N$-dimensional vector space. Moreover, $A^T=-A$ implies that it is antisymmetric. Then, $A$ must contain only $2\times 2$ blocks, like $\mathcal{R}_4(a)$ and $\mathcal{R}_4(a^3)$, with vanishing diagonal elements. So, for odd $N$, the $3N \times 3N$ matrix cannot be built using $2\times 2$ blocks.

In the case of $N$ even, it suffices to use Eq.\eqref{1.1.51} with $k=M$ and $r_i=0$, for $i\neq 2$. Indeed, Corollary \ref{cor:2.1} holds, and antisymmetry implies that the diagonal elements vanish, excluding cycles of order $1$. Moreover, it is trivial to make a one-to-one correspondence between $\mathcal{R}_4(a)$ and the representative of the non trivial element of $\mathbb{Z}_2$:
\begin{equation}
  \mathcal{R}(a)=
    \begin{pmatrix}
    0&1\\
    1&0
    \end{pmatrix}  
\end{equation}
Then, one must count the elements of a conjugation class built only with cycles of order $2$, which can be done using Eq.\eqref{1.1.51} and considering that the number of cycles of order $2$ over $M$ elements is $M/2$. The result is:
    \begin{equation}
    \Delta=\frac{M!}{2^{M/2}(M/2)!}
\end{equation}
Finally, unlike the case of Classical mechanics, in which $\mathcal{M}^2=I$, here we
also admit the opposite sign for the $2\times 2$ block, that corresponds to $\mathcal{R}_4(a^3)$. Multiplying $\Delta$ by $2^{M/2}$, we finally get Eq.\eqref{1.108}.
\end{proof}

\noindent
The difference between Classical and Quantum cases is easily revealed by the following example. The Hamiltonian of a system made of $2$ free quantum particles moving in $1$-dimension is given by $H=p_1^2/2m+p_2^2/2m$. Consider a time reversal operator $\mathcal{T}$ whose unitary part is represented on the vectors of  momentum operators $\boldsymbol P=(p_1,p_2)$ by the matrix
\begin{equation}
Q=
    \begin{pmatrix}
    0&-1\\
    1&0
    \end{pmatrix}
\end{equation}
While admitted in quantum mechanics, it is not acceptable in classical mechanics, because $Q^2=-I$. 

Having exhausted our treatment of spinless particles coupled with an external magnetic field, we now turn to the case of particles with spin $1/2$.

\subsection{Time reversal with spin}
\label{sec:Time reversal with spin}
For non-relativistic spin systems, the equation of motion is the Pauli equation:
\begin{equation}
    \label{eq:2.40}
    \left\{\sum_{j=1}^N{\frac {1}{2m_j}}[{\boldsymbol {\sigma }_j}\cdot (\boldsymbol {p}_j -q_j\boldsymbol{A}(x_j,y_j,z_j) )^{2} \right\}|\psi \rangle =i {\frac {\partial }{\partial t}}|\psi \rangle 
\end{equation}
where $\ket{\psi}$ represents the total state of the system of $N$ particles and the spin operator $\boldsymbol\sigma_j$ of the $j$-th particle is expressed by
\begin{equation}
\label{spacespin}
    \boldsymbol{\sigma}_j=I\otimes...\otimes\underbrace{\boldsymbol{\sigma}}_j\otimes...\otimes I
\end{equation} 
where $\boldsymbol\sigma=(\sigma_x,\sigma_y,\sigma_z)$ contains the Pauli matrices as Cartesian components and $I$ is the identity operator. The Pauli vector identity: 
\begin{equation}
\label{eq:2.41}
(\boldsymbol{\sigma}\cdot \boldsymbol{a} )({\boldsymbol{\sigma}}\cdot \boldsymbol{b} )=\boldsymbol{a} \cdot \boldsymbol{b} +i{\boldsymbol{\sigma }}\cdot(\boldsymbol{a} \times \boldsymbol{b})
\end{equation}
yields
\begin{equation}
    \label{eq:2.42}
    \sum_{j=1}^N{\frac {1}{2m_j}}\left[(\boldsymbol {p}_j -q_j\boldsymbol {A}(x_j,y_j,z_j) )^{2}-q_j\boldsymbol{\sigma}_j\cdot\boldsymbol{B}\right]|\psi \rangle =i{\frac {\partial }{\partial t}}|\psi \rangle 
\end{equation}
where $\boldsymbol{B}=\nabla\times\boldsymbol{A}$.
For the time reversal operator, we may adopt the following form:
\begin{equation}
    \mathcal{T}=(U_c\otimes U_s)K
\end{equation}
where $U_c$ denotes the part acting on the coordinate and momentum operators, as in Sec.\ref{sec:The time reversal operator}; while $U_s$ is the part acting on the spin operators. Both $U_c$ and $U_s$ must be unitary, in order to represent time reversal operators. Since $\boldsymbol\sigma_j$ separately acts on single particles spin operators, in the following we omit the subscript $j$, considering, without loss of generality, the effect of $\mathcal{T}$ on $\boldsymbol\sigma$. 

In this framework, time reversal has been 
traditionally defined as an operator $\mathcal{T}$ that preserves the commutation relations and,
analogously to classical mechanics, yields 
\cite{sachs}:
\begin{equation}
\label{1.95}
    \mathcal{T}\boldsymbol \sigma\mathcal{T}^{-1}=-\boldsymbol \sigma
\end{equation} 
\noindent
which directly implies:
\begin{equation}
    U_s=\sigma_y
\end{equation}
Unfortunately, this condition does not allow us to treat particles with spin \cite{rondoquantu}. On the other hand, Proposition \ref{prop:2.6} suggests we may relax some restriction. Therefore, we adopt the following, minimal, definition that does not include condition \eqref{1.95} and that actualizes Definition \ref{prop:1.2.1} to particles with spin 1/2. 

\begin{defi}[Particles with spin 1/2] 
\label{assu:2.4} 
An antilinear operator $\cal T$ verifying Definition \ref{prop:1.2.1} is called a time reversal operator for particles with spin, if it preserves the commutation relations of Pauli matrices:
\begin{equation}
\label{1.114}
    [\sigma_{j},\sigma_{k}]=i \tensor{\varepsilon} {_{jk}^{l}}\sigma_{l}
\end{equation}
where $j, k, l = x, y, z$ and $\tensor{\varepsilon} {_{jk}^{l}}$ is the Levi-Civita symbol.
\end{defi}

\noindent
Note that only $\sigma_y$ is complex and so
the complex conjugation operator $K$ preserves \eqref{1.114}. To proceed, let us recall two  results of Lie Group Theory \cite{gilmore2012lie},
taking $\mathbb{R}^3$ as a Lie algebra generated by three vectors $(\boldsymbol e_1,\boldsymbol e_2,\boldsymbol e_3)$:
\begin{theo}
\label{theo:2.4}
  There exists a linear invertible map $\Phi:\mathfrak{su}(2)\xrightarrow[]{}\mathbb{R}^3$ such that \begin{equation}
      \Phi\left(\left[\frac{i}{2}\sigma_j,\frac{i}{2}\sigma_k\right]\right)=\Phi\left(\frac{i}{2}\sigma_j\right)\times\Phi\left(\frac{i}{2}\sigma_k\right)\, ; \quad \mbox{with } ~~
      \Phi\left(\frac{i}{2}\sigma_j\right)=\boldsymbol e_j
  \end{equation}
where $\times$ (the usual cross product) is the Lie product.
\end{theo}
\begin{theo}
\label{theo:1.2.2}
An element of the unitary group $U\in U(2)$ can be expressed as 
\begin{equation}
\label{1.116}
    U=e^{\frac{i}{2}\lambda^0 }\left(I\cos{\frac{\lambda}{2}}+i\frac{\lambda^j\sigma_j}{\lambda}\sin{\frac{\lambda}{2}}\right)=e^{\frac{i}{2}\lambda^0 } V
\end{equation}
where $\lambda^j \in \mathbb{R}$ for $j=0,1,2,3$,
$\lambda=|\boldsymbol\lambda|$, for $\boldsymbol\lambda=(\lambda^1,\lambda^2,\lambda^3)$, and $V\in SU(2)$.
\end{theo}
\noindent
Theorem \ref{theo:2.4} allows us to find a $U_s$ that preserves the commutation relations \eqref{1.114}, by finding a transformation that preserves a right handed triad in $\mathbb{R}^3$. 
\begin{lemma}
\label{lemma1}
Linear transformations mapping a right handed basis in $\mathbb{R}^3$ into another belong to $SO(3)$.
\end{lemma}

\noindent
Another useful fact is that given $V\in SU(2)$ as in \eqref{1.116}, one has:
\begin{equation}
    \label{complex}
    \sigma_y U\sigma_y=U^{*}
\end{equation}
where $U^*$ is the complex conjugate of $U$. Then, thanks to the properties of the Pauli matrices, we obtain:
\begin{equation}
\begin{split}
    \sigma_y \left(I\cos{\frac{\lambda}{2}}+i\frac{\lambda^j\sigma_j}{\lambda}\sin{\frac{\lambda}{2}}\right)\sigma_y&=I\cos{\frac{\lambda}{2}}+i\frac{\lambda^1\sigma_y\sigma_x\sigma_y+\lambda^2\sigma_y\sigma_y\sigma_y+\lambda^3\sigma_y\sigma_z\sigma_y}{\lambda}\sin{\frac{\lambda}{2}}=\\&=\left(I\cos{\frac{\lambda}{2}}-i\frac{\lambda^j\sigma_j}{\lambda}\sin{\frac{\lambda}{2}}\right)
\end{split}
\end{equation}
Another useful classical result of group theory is the following:
\begin{theo}
\label{iso}
The quotient group $SU(2)/\mathbb{Z}_2$ is isomorphic to the group $SO(3)$. The isomorphism is defined by:
\begin{equation}
    U^\dagger \sigma_j U=[\Lambda(U)]_j^k\sigma_k
\end{equation}
where $\Lambda(U)$ is a $3\times 3$ special orthogonal matrix associated to $U\in SU(2)/\mathbb{Z}_2$ and we used Einstein summation rule. 
\end{theo}
For example, the operator $U_d$ acting on the spin operators of a single particle as
     \begin{equation}
  \label{1.117b}
    \begin{split}
   U_d\sigma_xU_d^{-1}=&s_x\sigma_x\\
   U_d\sigma_yU_d^{-1}=&s_y\sigma_y\\
    U_d\sigma_zU_d^{-1}=&s_z\sigma_z
    \end{split}
\end{equation}
can be associated with the matrix 
\begin{equation}
\Lambda(U_d)=
    \begin{pmatrix}
        s_x&0&0\\
        0&s_y&0\\
        0&0&s_z
    \end{pmatrix}
\end{equation}
and the whole equivalence class $\{U_d,-U_d\}$ is is represented by $\Lambda(U_d)$.

Let us now analyze the case in which $\Lambda(U)$ is a permutation and an involution, hence it is not a cyclic or counter-cyclic permutation of the basis vectors.
Proceeding analogously to the case of time reversal operations acting on coordinates, we obtain the following theorem concerning the spin space.

\begin{theo}
\label{theo:1.2.3}
Given an operator ${\cal T}=UK$, where $K$ is complex conjugation, a time reversal operator acting on the spin space is obtained if $U$ is unitary and obeys:
\begin{equation}
\label{1.119b}
    \begin{split}
       &U^{(1)}_{xy}=\theta(\sigma_z\mp iI)\\
       &U^{(1)}_{yz}=\theta(\sigma_x\mp i I)\\
       &U^{(2)}_{xz}=\theta(\sigma_x\pm\sigma_z)\\
       &U_d=\sigma_j
    \end{split}
\end{equation}
where $\theta\in\mathbb{C}$, $|\theta|=1/\sqrt{2}$, and $j=x,y,z$.
\end{theo}
\begin{proof}
  First, we consider the diagonal transformations $U_d$ that do not permute the Pauli matrices:
  \begin{equation}
  \label{1.117}
    \begin{split}
   U_d\sigma_xU_d^{-1}=&s_x\sigma_x\\
   U_d\sigma_yU_d^{-1}=&s_y\sigma_y\\
    U_d\sigma_zU_d^{-1}=&s_z\sigma_z
    \end{split}
\end{equation}
where $s_i=\pm1$, $i=x,y,z$, to ensure the involution nature of the operation. Moreover, since the triad must remain right handed we need that $s_xs_ys_z=1$. Then, apart from the trivial case $s_x=s_y=s_z=1$, there are three possibilities. First, the case $s_y=1$ and $s_x=s_z=-1$ that corresponds to the canonical situation, since Eqs.\eqref{1.117} become 
\begin{equation}
  \label{1.118}
    \begin{split}
   U_d\sigma_xU_d^{-1}=&-\sigma_x\\
   U_d\sigma_yU_d^{-1}=&\sigma_y\\
    U_d\sigma_zU_d^{-1}=&-\sigma_z
    \end{split}
\end{equation}
whose solution is $U_d=\sigma_y$. The second case is $s_x=1$ and $s_y=s_z=-1$, which leads to 
$U_d=\sigma_x$. The third case, $U_d=\sigma_z$, holds for the $s_z=1$ and $s_x=s_y=-1$. 
This leads to $\mathcal{T}^2=U_d K U_d K= \pm I$ as requested. 
Regarding $U_d=\sigma_x$ and $U_d=\sigma_z$ the action of $K$ is irrelevant since these matrices are real, hence:
\begin{equation}
    \mathcal{T}^2=\sigma_x^2=\sigma_z^2 = I
\end{equation}
Lastly, the case $U_d=\sigma_y$, gives $\mathcal{T}^2=-\sigma_y^2= - I$.

Concerning transformations that permute the order of the basis elements, and so of the Pauli matrices, we have the following possibilities: if we want to permute $\sigma_x$ and $\sigma_y$ we  take either:
\begin{equation}
    \label{1.122}
    \begin{split}
    U^{(1)}_{xy}\sigma_x(U^{(1)}_{xy})^{-1}=&\mp\sigma_y\\
    U^{(1)}_{xy}\sigma_y(U^{(1)}_{xy})^{-1}=&\pm\sigma_x\\
    U^{(1)}_{xy}\sigma_z(U^{(1)}_{xy})^{-1}=&\sigma_z
    \end{split}
\end{equation}
or
\begin{equation}
    \label{1.121}
    \begin{split}
    U^{(2)}_{xy}\sigma_x(U^{(2)}_{xy})^{-1}=&\pm\sigma_y\\
    U^{(2)}_{xy}\sigma_y(U^{(2)}_{xy})^{-1}=&\pm\sigma_x\\
    U^{(2)}_{xy}\sigma_z(U^{(2)}_{xy})^{-1}=&-\sigma_z
    \end{split}
\end{equation}
To permute $\sigma_y$ and $\sigma_z$  we  take either:
\begin{equation}
    \label{eq:spin19}
   \begin{split}
    U^{(1)}_{yz}\sigma_x(U^{(1)}_{yz})^{-1}=&\sigma_x\\
    U^{(1)}_{yz}\sigma_y(U^{(1)}_{yz})^{-1}=&\mp\sigma_z\\
    U^{(1)}_{yz}\sigma_z(U^{(1)}_{yz})^{-1}=&\pm\sigma_y
    \end{split}
\end{equation}
or
\begin{equation}
    \label{eq:spin20}
    \begin{split}
   U^{(2)}_{yz}\sigma_x(U^{(2)}_{yz})^{-1}=&-\sigma_x\\
   U^{(2)}_{yz}\sigma_y(U^{(2)}_{yz})^{-1}=&\pm\sigma_z\\
    U^{(2)}_{yz}\sigma_z(U^{(2)}_{yz})^{-1}=&\pm\sigma_y
    \end{split}
\end{equation}
To permute $\sigma_x$ and $\sigma_z$  we  take either:
\begin{equation}
    \label{eq:spin23}
    \begin{split}
    U^{(1)}_{xz}\sigma_x(U^{(1)}_{xz})^{-1}=&\mp\sigma_z\\
    U^{(1)}_{xz}\sigma_y(U^{(1)}_{xz})^{-1}=&\sigma_y\\
    U^{(1)}_{xz}\sigma_z(U^{(1)}_{xz})^{-1}=&\pm\sigma_x
    \end{split}
\end{equation}
or
\begin{equation}
    \label{eq:spin24}
    \begin{split}
   U^{(2)}_{xz}\sigma_x(U^{(1)}_{xz})^{-1}=&\pm\sigma_z\\
   U^{(2)}_{xz}\sigma_y(U^{(1)}_{xz})^{-1}=&-\sigma_y\\
    U^{(2)}_{xz}\sigma_z(U^{(1)}_{xz})^{-1}=&\pm\sigma_x
    \end{split}
\end{equation}
Here, we required that the triad remains right handed, so the representation on the associated vectors $\boldsymbol e_i$ (see Theorem \ref{theo:2.4}) is a matrix of $SO(3)$. For instance, Eqs.\eqref{1.122} are associated with a matrix
\begin{equation}
    O=\begin{pmatrix}
    0&\mp 1&0\\
    \pm 1&0&0\\
    0&0&1
    \end{pmatrix}
\end{equation}
that is special orthogonal. In fact, it is the linear application of $SO(3)$ that maps $(\boldsymbol e_x,\boldsymbol e_y,\boldsymbol e_z)$ to $(\mp\boldsymbol e_y,\pm\boldsymbol e_x,\boldsymbol e_z)$.
Now, use Theorem \ref{theo:1.2.2}. Because $U_{xy}$ must be unitary, we express it as the following linear combination:
\begin{equation}
\label{1.123}
    U_{xy}=\alpha\sigma_x+\beta\sigma_y+\gamma\sigma_z+\delta I\quad\quad \alpha,\beta,\gamma,\delta\in\mathbb{C}
\end{equation}
whose coefficients obey Eq. \eqref{1.116}.  
Substituting \eqref{1.123} in the third equation of Eqs.\eqref{1.122}, we obtain the condition:
\begin{equation}
\label{1.124}
    \alpha\sigma_x\sigma_z+\beta\sigma_y\sigma_z+\gamma I+\delta \sigma_z=\alpha\sigma_z\sigma_x+\beta\sigma_z\sigma_y+\gamma I+\delta \sigma_z
\end{equation}
and, recalling that the Pauli matrices satisfy the fundamental relation 
\begin{equation}
\label{1.145}
    \sigma_j\sigma_k=\delta_{jk}I+i\tensor{\varepsilon} {_{jk}^{l}}\sigma_l
\end{equation}
Eq\eqref{1.124} writes 
\begin{equation}
\label{line}
    -i\alpha\sigma_y+i\beta\sigma_x+\gamma I+\delta\sigma_z=i\alpha\sigma_y-i\beta\sigma_x+\gamma I+\delta\sigma_z
\end{equation}
Since the set $\{\sigma_j,I\}$ is a basis for the algebra $\mathfrak{u}(2)$, the coefficient of the linear combinations on both sides of Eq.\eqref{line} must coincide, which means $\alpha=\beta=0$, and $U_{xy}=\gamma\sigma_z+\delta I$. Substituting in the first of Eqs.\eqref{1.122} one gets:
\begin{equation}
\label{1.125}
 \gamma \sigma_z\sigma_x+\delta \sigma_x=\mp\gamma \sigma_y\sigma_z\mp\delta \sigma_y
\end{equation}
that is 
\begin{equation}
\label{1.126}
 i\gamma \sigma_y+\delta \sigma_x=\mp i\gamma \sigma_x\mp\delta \sigma_y
\end{equation}
which implies $\delta=\mp i\gamma$. To ensure that the corresponding matrix 
\begin{equation}
    U^{(1)}_{xy}=\gamma\sigma_z\mp i\gamma I=\gamma(\sigma_z\mp iI)
\end{equation}
is unitary, we need $UU^\dagger=I$, that is: 
\begin{equation}
    \gamma^2(\sigma_z\mp iI)(\sigma_z\pm iI)=2\gamma^2 I=I 
\end{equation}
which means: $|\gamma|=1/\sqrt{2}$. The last thing to check, for a time reversal operator,
is that it is an involution, {\em i.e.}\ that $\mathcal{T}^2=UKUK=\pm I$ holds. This is indeed the case:
\begin{equation}
    U^{(1)}_{xy} K U^{(1)}_{xy} K=\gamma^2(\sigma_z\mp iI)(\sigma_z\pm i I)=I
\end{equation}
The same reasoning can be repeated for the remaining permutations, which leads to:
\begin{equation}
    \begin{split}
        U^{(2)}_{xy}=\theta(\sigma_x\pm\sigma_y)\\
        U^{(1)}_{yz}=\theta(\sigma_x\mp i I)\\
         U^{(2)}_{yz}=\theta(\sigma_y\pm\sigma_z)\\
         U^{(1)}_{xz}=\theta(\sigma_y\pm i I)\\
         U^{(2)}_{xz}=\theta(\sigma_x\pm\sigma_z)\\
    \end{split}
\end{equation}
where $\theta$ is a complex number with modulus $|\theta|=1/\sqrt{2}$. One may easily check that $\mathcal{T}^2=\pm I$ in all cases:
\begin{equation}
    \begin{split}
     U^{(2)}_{xy} K U^{(2)}_{xy} K=\theta^2(\sigma_x\pm\sigma_y)(\sigma_x\mp\sigma_y)=\mp i\sigma_z\neq\pm I\\
     U^{(1)}_{yz} K U^{(1)}_{yz} K=\theta^2(\sigma_x\mp i I)(\sigma_x\pm iI)=I\\
     U^{(2)}_{yz} K U^{(2)}_{yz} K=\theta^2(\sigma_y\pm \sigma_z)(-\sigma_y\pm\sigma_z)=\pm i\sigma_x\neq\pm I\\
      U^{(1)}_{xz} K U^{(1)}_{xz} K=\theta^2(\sigma_y\pm i I)(-\sigma_y\mp i I)=\mp 2i\sigma_y\neq\pm I\\
       U^{(2)}_{xz} K U^{(2)}_{xz} K=\theta^2(\sigma_x\pm\sigma_z)(\sigma_x\pm\sigma_z)=I
    \end{split}
\end{equation}
The Theorem is proven, taking $U^{(1)}_{xy}$, $U^{(1)}_{yz}$ and $U^{(2)}_{xz}$ as the non diagonal time reversal operators on spin space. 
\end{proof}
Now, swapping a pair of coordinates requires at most $2$ matrices. The total, considering also $U_d$, makes $9$ possible matrices $U_s$. Because, each of them is defined up to a complex number of modulus $1$, it seems that infinitely many are allowed. But whenever $U_s$ is applied to an operator on the spin space, giving $U_s f(\sigma_j) U_s^{-1}$, that number is eliminated. Therefore, there are $9$ operators acting differently on Pauli matrices.

This exhausts our present investigation of generalized time reversal invariance, for quantum non-relativistic systems, with and without half integer spin. In the following,
we apply our results to the theory of the Onsager relations, looking for generalizations of the compatibility conditions found in Ref.\cite{carbone2020necessary}, between time reversal transformations and a generic magnetic field.


\subsection{Compatibility condition between time reversal operations and magnetic field}
\label{sec:compatib}
The reasoning developed by Kubo about correlation functions and the entailing transport coefficients \cite{kubo1957statistical}, requires the existence of a time reversal operation, that commutes with the Hamiltonian of the system of interest. Because the quantum observables are defined as hermitian operators, one cannot follow the classical procedure to derive correlator relations. Given two observables $\phi$ and $\psi$, the mean value $\langle\phi(0)\psi(t)\rangle$ is ill-defined, as the combined operator $\phi(0)\psi(t)$ is not guaranteed to be hermitian. Therefore, Kubo introduced the \textit{canonical} correlator in Response Theory as:
\begin{equation}
    \label{eq:2.23}
    \langle\phi(0);\psi(t)\rangle=\frac{1}{\beta}\int_0^\beta d\lambda\;\; \mathbf{Tr}[\rho e^{\lambda H}\phi(0)e^{-\lambda H}\psi(t)]
\end{equation}
where $\rho$ is the density operator, $H$ the hamiltonian, and {\bf Tr} the trace over the  Hilbert. Kubo then proved that this is indeed
a real quantity.  Note that non-relativistic systems with spin are allowed: it suffices to include the spin degrees of freedom within the Hilbert space. Another possible definition is to symmetrize the correlator as $\langle\phi(0),\psi(t)\rangle=\langle\phi(0)\psi(t)\rangle+\langle\psi(t)\phi(0)\rangle$, but Kubo showed that \ref{eq:2.23} is sufficiently general.
The sufficient condition for the validity of the Onsager relations can then be formulated as {\em e.g.}\ Ref.\cite{rondoquantu}:
\begin{prop}
\label{prop:2.3}
Consider a quantum mechanical particle system in a magnetic field $\bf B$, with Hamiltonian $H$ and density matrix $\rho=\rho(H)$. Let $\mathcal{T}=UK$ be a time reversal operator that commutes with $H$, and let $\phi$ and $\psi$ be two observables with 
signatures $\eta_\phi$ and $\eta_\psi$ with respect to $\cal T$, {\em i.e}\
\begin{equation}
    \label{eq:2.24}
    \mathcal{T}\phi\mathcal{T}^{-1}=\eta_\phi\phi\;\;\;\;\;\;\mathcal{T}\psi\mathcal{T}^{-1}=\eta_\psi\psi
\end{equation}
Then, the equality:
\begin{equation}
    \langle\phi(0);\psi(t)\rangle_{\bf B} =
    \eta_\phi \eta_\psi \langle\phi(t);\psi(0)\rangle_{\bf B}
    \label{PhiPsiB}
\end{equation}
holds.
\end{prop}
\noindent
\begin{rem}
Assuming that the density matrix depends on $H$ is quite general; for example, the canonical ensemble satisfies this constraint. Moreover, this assumption is sufficient for deduce the results of Proposition \eqref{prop:2.6}; it is not necessary to clarify what kind of statistical ensemble (bosons, fermions, anyons etc.) we are considering.
\end{rem}
This is the core of the Onsager reciprocal relations in Quantum mechanics in the presence of a magnetic field. Therefore, we now investigate how particular magnetic fields affect the number of time reversal operations consistent with Eq.\eqref{PhiPsiB}.
Consider the following Hamiltonian for a system of spinless particles coupled with a potential vector $\mathbf{A}$:
\begin{equation}
\label{eq:2.27}
   H=\sum_{i=1}^N\left[\frac{(\mathbf{p}_i-q_i\mathbf{A}(x_i,y_i,z_i))^2}{2m_i}\right]
\end{equation}
where $N$ is the number of particles, $q_i$ and $m_i$ are the charge and the mass of the $i$-th particle, and
\begin{equation}
\label{eq:2.28}
    (\mathbf{p}-q\mathbf{A})^2=-\nabla^2+iq\nabla\cdot\mathbf{A}+iq\mathbf{A}\cdot\nabla+q^2\mathbf{A}^2
\end{equation}

The commonly used Coulomb gauge $\nabla\cdot\boldsymbol{A}=0$ that importantly effects on this expression eliminating the second addend. As well-known, the vector potential associated to a magnetic field is defined up to the gradient of a scalar function since $\boldsymbol B=\nabla \times \boldsymbol A$. Thus, we can define an equivalence class containing the vector potentials that originate the same magnetic field.  In the following, we refer to $[\boldsymbol{A}(\boldsymbol{x})]_R$ to denote a representative of the class containing $\boldsymbol{A}(\boldsymbol{x})$.

After this clarifications, let us step back to the classical case: we now show the core theorem that regards the compatibility between a particular time reversal operation and a generic magnetic field (and so vector potential). We point out as in the following we are going to refer to $V(\mathbb{R}^{n})$ as the space of vector fields on $\mathbb{R}^{n}$.
\begin{theo}[Compatibility conditions for $\boldsymbol A$ in Classical Mechanics]
\label{theo:2.7}
Consider a system of $N$ particles of equal mass $m$ and charge $q$. Let $\mathcal{M}$ be a generalized time reversal operator acting on the coordinates as $({\cal M} A)(\boldsymbol X)=A({\cal M}_M \boldsymbol X)$, where $\mathcal{M}_M\in O(3N)$. Denote by $\boldsymbol{\mathcal{A}}(\boldsymbol X)=(\boldsymbol{A}(\boldsymbol{x}_1),...,\boldsymbol{A}(\boldsymbol{x}_N))$ the 
$3N$ dimensional vectors of coordinates, and by $[\boldsymbol{\mathcal{A}}(\boldsymbol X)]_R=([\boldsymbol{A}(\boldsymbol{x}_1)]_R,...,[\boldsymbol{A}(\boldsymbol{x}_N)]_R)$ the corresponding equivalence class. Introduce the reversal operator $\mathcal{M}':V(\mathbb{R}^{3N})\xrightarrow{}V(\mathbb{R}^{3N})$ defined by: 
\begin{equation}
\label{3.13bb}
    (\mathcal{M}'\boldsymbol{\mathcal{A}})(\boldsymbol X)\equiv\mathcal{M}_M\boldsymbol{\mathcal{A}}(\mathcal{M}_M\boldsymbol X)
\end{equation}
The operator $\mathcal{M}$ yields TRI in the presence of a magnetic vector potential $\boldsymbol{A}$, if and only if
\begin{equation}
\label{3.13}
    (\mathcal{M}'\boldsymbol{\mathcal{A}})(\boldsymbol X)=[-\boldsymbol{\mathcal{A}}(\boldsymbol X)]_R
\end{equation} 
\end{theo}

\begin{proof}
Recall that TRI holds if there is a time reversal transformation that preserves the equations of motion, as well as the Hamiltonian up to a gauge transformation.
In our case, the equations of motion are invariant under a gauge transformation, as can be easily seen. Then, in the presence of the minimal coupling with a magnetic field, the condition $H(\mathcal{M}\Gamma)=H(\Gamma)$ is verified for a gauge transformation, for any $\Gamma$. Therefore, the Hamiltonian can be written as
\begin{equation}
\label{3.14}
   H(\boldsymbol X,\boldsymbol P)=\frac{1}{2m}\sum_i[\boldsymbol{p}_i -q\boldsymbol{A}(\boldsymbol{x}_i) ]^{2}=\frac{1}{2m}[\boldsymbol{P} -q\boldsymbol{\mathcal{A}}(\boldsymbol{X})]^2  = \frac{1}{2m}[\boldsymbol{P}+ q\mathcal{M}_M\boldsymbol{\mathcal{A}}(\mathcal{M}_M\boldsymbol{X})] \cdot [\boldsymbol{P}+ q\mathcal{M}_M\boldsymbol{\mathcal{A}}(\mathcal{M}_M\boldsymbol{X})]
\end{equation}
where $\boldsymbol{P}$ and $\mathcal{A}$ are  $3N$-dimensional vectors. By Theorem \ref{theo:1.1.2}, a time reversal operation for systems subject to a magnetic field must act separately on coordinates and momenta, namely the transformation must be expressed by: 
\begin{equation}
\label{3.15}
    (\boldsymbol X, \boldsymbol P)\xrightarrow[]{\mathcal{M}}(\mathcal{M}_M\boldsymbol X,-\mathcal{M}_M\boldsymbol P) 
\end{equation}
with $\mathcal{M}_M\in O(3N)$. This yields
\begin{equation}
\label{3.16}
\begin{split}
       H(\mathcal{M}_M\boldsymbol X,-\mathcal{M}_M\boldsymbol P)&=\frac{1}{2m}[-\mathcal{M}_M\boldsymbol{P} -q\boldsymbol{\mathcal{A}}(\mathcal{M}_M\boldsymbol{X})]^2\\
       &=\frac{1}{2m}[\boldsymbol{P}+ q\mathcal{M}_M\boldsymbol{\mathcal{A}}(\mathcal{M}_M\boldsymbol{X})]^2
\end{split}
\end{equation}
where we used the fact that the scalar product is invariant under rotations such as $\mathcal{M}_M$. 
Now, let $\cal G$ be the scalar function involved in the gauge choice, define $\nabla_{3N}\mathcal{G}(\boldsymbol X)=(\nabla G(\boldsymbol{x}_1),...,\nabla G(\boldsymbol{x}_N ))$, and let us introduce the equivalence class of vector potentials defined by:
\begin{equation}
\label{3.17}
    [\boldsymbol{A}]=\{\boldsymbol{A}\;\;|\;\; \nabla\times \boldsymbol A=\boldsymbol B\}
\end{equation}
This is the set of vector potentials corresponding to a given magnetic field $\boldsymbol B$. One representative element of this class, denoted by a subscript $R$, is expressed by: 
\begin{equation}
    [\boldsymbol{\mathcal{A}}(\boldsymbol X)]_R=([\boldsymbol{A}(\boldsymbol{x}_1)]_R,...,[\boldsymbol{A}(\boldsymbol{x}_N)]_R)
\end{equation}
Now, taking
$\mathcal{M}_M\boldsymbol{\mathcal{A}}(\mathcal{M}_M\boldsymbol{X})=-(\boldsymbol{\mathcal{A}}(\boldsymbol{X})+\nabla_{3N}\mathcal{G}(\boldsymbol X))$, substitution shows that
\eqref{3.14} and \eqref{3.16} are equal, up to a gauge transformation. The transformation leaves the equations of motion unchanged. 
If, on the other hand, \eqref{3.14} and \eqref{3.16} are equivalent for any value of $\boldsymbol P$ and $\boldsymbol X$, that is:
\begin{equation}
\label{3.19}
\begin{split}
    &\hskip -50pt
    \boldsymbol{P}^2+2q \boldsymbol{P}\cdot(\boldsymbol{\mathcal{A}}(\boldsymbol{X})+\nabla_{3N}\mathcal{G}(\boldsymbol X))+q^2(\boldsymbol{\mathcal{A}}(\boldsymbol{X})+\nabla_{3N}\mathcal{G}(\boldsymbol X))^2\\
    &\hskip 50pt =\boldsymbol{P}^2-2q\boldsymbol{P}\cdot\mathcal{M}_M\boldsymbol{\mathcal{A}}(\mathcal{M}_M\boldsymbol{X})+q^2(\boldsymbol{\mathcal{M}_M\mathcal{A}}(\mathcal{M}_M\boldsymbol{X}))^2 \, , \quad \forall \boldsymbol P
\end{split}
\end{equation}
we can choose $\boldsymbol P=0$ obtaining
\begin{equation}
    q^2(\boldsymbol{\mathcal{A}}(\boldsymbol{X})+\nabla_{3N}\mathcal{G}(\boldsymbol X))^2=q^2(\boldsymbol{\mathcal{M}_M\mathcal{A}}(\mathcal{M}_M\boldsymbol{X}))^2
\end{equation}
Then, Eq.\eqref{3.19} reduces to
\begin{equation}
\label{3.20}
    2q \boldsymbol{P}\cdot(\boldsymbol{\mathcal{A}}(\boldsymbol{X})+\nabla_{3N}\mathcal{G}(\boldsymbol X))=-2q\boldsymbol{P}\cdot\mathcal{M}_M\boldsymbol{\mathcal{A}}(\mathcal{M}_M\boldsymbol{X})
\end{equation}
Therefore, $\mathcal{M}_M\boldsymbol{\mathcal{A}}(\mathcal{M}_M\boldsymbol{X})=-(\boldsymbol{\mathcal{A}}(\boldsymbol{X})+\nabla_{3N}\mathcal{G}(\boldsymbol X))$ holds.
\end{proof}
\noindent
This result generalizes the compatibility conditions of Ref.\cite{carbone2020necessary}, which focused on the $20$ transformations that separately act on each particle subspace. The result of Ref.\cite{carbone2020necessary} can be interpreted as a corollary of Theorem \ref{theo:2.7}.
\begin{cor}
Take a block diagonal transformation $\mathcal{M}_M$ separately acting on each particle subspace, denoting by $\mathcal{M}_m$ one such $3\times3$ block. Then
\begin{equation}
\label{3.21}
    \mathcal{M}'\boldsymbol{A}=\mathcal{M}_m\boldsymbol{A}(\mathcal{M}_m\boldsymbol X)=[-\boldsymbol{A}]_R
\end{equation}
\end{cor}
\begin{proof}
It trivially follows form the definition of $\boldsymbol{\mathcal{A}}$ given in Theorem \ref{theo:2.7}, and from Eq.\eqref{3.13}, where $\mathcal{M}_M$ is a block diagonal acting on a single $3$-dimensional particle subspace.
\end{proof}

\begin{prop}
\label{prop:1.1.7}
Take the Hamiltonian \eqref{eq:2.27} in which $(m_i,q_i) \ne (m_j,q_j)$ if $i \ne j$ are particle indices. Then, a time reversal operator yielding TRI is given by a block diagonal matrix $A$, whose blocks separately act on single particle subspaces.   
\end{prop}
\begin{proof}
By {\em reductio ad absurdum}, suppose a valid non diagonal time reversal operation $\mathcal{M}_{nd}$ exists for the system  described by the Hamiltonian \eqref{eq:2.27}. Without loss of generality, assume that the operation acts on the $M \times M$ coordinates (with $M=3N$) swapping those of particle $j$ with those of particle $j+1$:
\begin{equation}
\label{1.1.39b}
     (\boldsymbol{x}_1,...,\boldsymbol{x}_j,\boldsymbol{x}_{j+1},...,\boldsymbol{x}_N,\boldsymbol{p}_1,...,\boldsymbol{p}_j,\boldsymbol{p}_{j+1},...,\boldsymbol{p}_N)
     \xrightarrow{\mathcal{M}_{nd}}(\boldsymbol{x}_1,...,\boldsymbol{x}_{j+1},\boldsymbol{x}_j,...,\boldsymbol{x}_N,
     -\boldsymbol{p}_1,...,-\boldsymbol{p}_{j+1},-\boldsymbol{p}_{j},...,-\boldsymbol{p}_N )
\end{equation}
 By hypothesis,  $H(\mathcal{M}_{nd}\Gamma)=H(\Gamma)$, because $\mathcal{M}_{nd}$ yields TRI. It suffices to check the contributions due to particles $j$ and $j+1$, because all other contributions are unchanged. Thus we can write:
\begin{eqnarray}
    \label{1.1.40}
    \sum_{i=1}^N\left[\frac{(\boldsymbol{p}_i-q_i\boldsymbol{A}(\boldsymbol{x}_i))^2}{2m_i}\right] &=& ...+\frac{(\boldsymbol{p}_j-q_j\boldsymbol{A}(\boldsymbol x_j))^2}{2m_j}+\frac{(\boldsymbol{p}_{j+1}-q_{j+1}\boldsymbol{A}(\boldsymbol x_{j+1}))^2}{2m_{j+1}}+...
\\
    \label{1.1.41}
    &&...+\frac{(\boldsymbol{p}_{j+1}+q_j\boldsymbol{A}(\boldsymbol x_{j+1}))^2}{2m_{j}}+\frac{(\boldsymbol{p}_{j}+q_{j+1}\boldsymbol{A}(\boldsymbol x_{j}))^2}{2m_{j+1}}+...
\end{eqnarray}
But $m_i\neq m_j$ or $q_i\neq q_j$ for $\forall i\neq j$ by hypothesis. Then, in general, one has $H(\mathcal{M}_{nd}\Gamma)\neq H(\Gamma)$, unless special cases are considered, because the coordinates and the momenta may take any value in $\mathbb{R}^3$. This is absurd. As the reasoning can be repeated varying $i$ and $j$ in \eqref{1.1.39b}, $\mathcal{M}_{nd}$ cannot mix the different single particles coordinates and momenta; it must be block diagonal, with each block acting on a single particle subspace.
\end{proof}

\noindent
This result may seem to be a drastic limitation on the range of TRI, but in reality it is not. Systems with particles of same charge and mass are the most widely  studied, both theoretically and experimentally.

\vskip 5pt

In Ref.\cite{carbone2020necessary}, the compatibility conditions is expressed also in terms of the magnetic field, instead of the vector potential. Here, we present an alternative derivation of that result result.  We point out as in the following we are going to refer to $V_p(\mathbb{R}^{n})$ as the space of pseudovector fields on $\mathbb{R}^{n}$.
\begin{theo}[Restricted compatibility conditions for $\boldsymbol B$ in Classical Mechanics]
\label{theo:3.1.2}
Consider a system of $N$ particles of equal mass $m$ and charge $q$. Let $\mathcal{M}$ be a generalized time reversal operator acting on the coordinates as $\mathcal{M}\boldsymbol X=(\mathcal {M}_m \boldsymbol x_1,...,\mathcal {M}_m \boldsymbol x_N)$, where $\mathcal{M}_m\in O(3)$. Introduce the pseudovector field rotation operator $\mathcal{M}':V_p(\mathbb{R}^{3})\xrightarrow{}V_p(\mathbb{R}^{3})$ defined by: 
\begin{equation}
\label{3.13c}
    (\mathcal{M}'\boldsymbol{B})(\boldsymbol x)=\operatorname{det}(\mathcal{M}_m)\mathcal{M}_m\boldsymbol{B}(\mathcal{M}_m\boldsymbol x)
\end{equation}
The operator $\mathcal{M}$ yields TRI in the presence of a magnetic field $\boldsymbol{B}$, if and only if
\begin{equation}
\label{3.22}
    (\mathcal{M}'\boldsymbol{B})(\boldsymbol x)=-\boldsymbol{B}(\boldsymbol x)
\end{equation} 
\end{theo}
\begin{proof}
We have to prove that \eqref{3.21} and \eqref{3.22} are equivalent. Let us introduce the notation:
\be
 \nabla_{\boldsymbol x}
 \equiv(\partial_x,\partial_y,\partial_z) 
\ee
hence $\nabla_{{\cal M}_m \boldsymbol x}$ is the gradient with respect to the coordinates rotated by ${\cal M}_m$. Then, keeping in mind that $\boldsymbol B = \nabla\times \boldsymbol A$, the application of the transformation $\mathcal{M}'$ defined by the orthogonal matrix $\mathcal{M}_m$ yields \cite{milne1948vectorial}: 
\begin{equation}
\label{3.23}
\begin{split}
     \mathcal{M}'(\nabla_{\boldsymbol x} \times \boldsymbol A)&=(\mathcal{M}_m\nabla_{\mathcal{M}_m\boldsymbol x})\times(\mathcal{M}_m\boldsymbol A \circ \mathcal{M}_m\boldsymbol)\\
     &=\operatorname{det}(\mathcal{M}_m)\mathcal{M}_m(\nabla_{\mathcal{M}_m\boldsymbol x} \times \boldsymbol A \circ \mathcal{M}_m)\\
     &=\operatorname{det}(\mathcal{M}_m)\mathcal{M}_m\boldsymbol B \circ \mathcal{M}_m
\end{split}
\end{equation}
because an orthogonal matrix such as $\mathcal{M}_m$ yields:
\begin{equation}
\nabla_{\mathcal{M}_m\boldsymbol x} \times \boldsymbol A \circ \mathcal{M}_m = \nabla\times \boldsymbol A = {\boldsymbol B} \, , 
\quad
    \mathcal{M}_m\nabla_{\mathcal{M}_m\boldsymbol x}=\nabla_{\boldsymbol x}
\end{equation}
Now, assuming Eq.\eqref{3.21} holds, we have:
\begin{equation}
    \operatorname{det}(\mathcal{M}_m)\mathcal{M}_m\boldsymbol B \circ \mathcal{M}_m =\nabla_{\boldsymbol x}\times[-\boldsymbol{A} ]_R=-\boldsymbol B
\end{equation}
Vice versa, assuming Eq.\eqref{3.22} holds, we start from 
\begin{equation}
    \operatorname{det}(\mathcal{M}_m)\mathcal{M}_m \nabla_{\mathcal{M}_m\boldsymbol x}\times \boldsymbol A \circ \mathcal{M}_m =-\nabla_{\boldsymbol x} \times \boldsymbol A
\end{equation}
Moreover, using the first and second lines of \eqref{3.23}, we get
\begin{equation}
   \nabla_{\boldsymbol x} \times (\mathcal{M}_m\boldsymbol A \circ \mathcal{M}_m) =-\nabla_{\boldsymbol x} \times \boldsymbol A
\end{equation}
and so, considering the fact that the rotor of a gradient is null, we obtain the thesis \eqref{3.21}, where gauge freedom is included.
\end{proof}
In the presence of magnetic fields, Theorem \ref{theo:2.7} seems to directly apply only in the case of the block diagonal time reversal operations.
Unlike Theorem \ref{theo:2.7},
Theorem \ref{theo:3.1.2} cannot be immediately generalised to the cases with a magnetic field, because in $\mathbb{R}^{3N}$ we miss the analogue of the relation $\boldsymbol A=\nabla\times\boldsymbol B$, which holds in $\mathbb{R}^3$. Operators similar to the curl can be defined in spaces other than $\mathbb{R}^3$, see for example Ref.\cite{conlon2008differentiable}, but we cannot claim that $\boldsymbol{\mathcal{A}}=\nabla \times \mathcal{B}$, for  $\boldsymbol{\mathcal{A}} , \boldsymbol{\mathcal{B}}\in\mathbb{R}^{3N}$.
Finally coming to the quantum context, the following holds.

\begin{prop}
\label{prop:3.1.2}
The statements of Theorems \ref{theo:2.7} and \ref{theo:3.1.2} extend to Quantum Mechanics on compact spaces, when $\boldsymbol X$ is identified with the position operator, cf.\ Theorem \ref{theo:1.2.1}.
\end{prop}

\begin{proof}
Let us start from Theorem \ref{theo:2.7}: regarding \eqref{3.14} and \eqref{3.15} the only differences are that $\boldsymbol P=-i\nabla$ and that the case $\mathcal{M}_M^2=-I$ is admitted. The proof can be repeated until \eqref{3.19}, since using Coulomb gauge and \eqref{eq:2.28} we do not have the $2$ in front of the mixed product. As a side note, the Coulomb gauge condition involves a scalar product too and so it invariant under rotation. \\
At this point, since we are using operators, we cannot impose $\boldsymbol P=0$ as done in Classical Mechanics. Nevertheless, \eqref{3.19} is an equivalence between operators and so it must holds for any wave function defined on a certain domain. If the domain is compact, we can consider the case of the constant without issues of normalization; the application to it of the differential operator $\boldsymbol P$ yields the constraint \begin{equation}
    q^2(\boldsymbol{\mathcal{A}}(\boldsymbol{X})+\nabla_{3N}\mathcal{G}(\boldsymbol X))^2=q^2(\boldsymbol{\mathcal{M}_M\mathcal{A}}(\mathcal{M}_M\boldsymbol{X}))^2
\end{equation}
Finally, we obtain an expression similar to \eqref{3.20}:
\begin{equation}
    (\boldsymbol{\mathcal{A}}(\boldsymbol{X})+\nabla_{3N}\mathcal{G}(\boldsymbol X))\cdot \boldsymbol{P}=-\mathcal{M}_M\boldsymbol{\mathcal{A}}(\mathcal{M}_M\boldsymbol{X})\cdot\boldsymbol{P}
\end{equation}
where we used the correct ordering of the momentum operators. Also in this case the only way to verify this operator equality is to have $\mathcal{M}_M\boldsymbol{\mathcal{A}}(\mathcal{M}_M\boldsymbol X)=[-\boldsymbol{\mathcal{A}}]_R$.\\
Coming to Theorem \ref{theo:3.1.2}, its extension is trivial since the proof involves only functions of the coordinate operator which acts in a multiplicative way in coordinate representation; in this way, we can proceed as if we were manipulating numbers and not operators, easily reproducing the demonstration. 
\end{proof}
This is our main result for the spinless case. Two observations and a fundamental example are in order.
\begin{rem}
\normalfont
The compatibility condition on magnetic fields of Theorem \ref{theo:3.1.2} seems to be of limited applicability, compared to the one about vector potentials, since it is limited to transformations separately acting on single particle subspaces. Nevertheless, that is the most interesting case, since particles usually interact with each other. For instance, in the case of the central interaction potential:
\begin{equation}
\label{poten}
    V=\sum_{i<j}v(\boldsymbol x_i-\boldsymbol x_j)
\end{equation}
with given pair potential $v(\boldsymbol x)=v(|\boldsymbol x|)$, treated in Ref.\cite{carbone2020necessary}, the time reversal operations must be block diagonal. It must separately and identically act on each particle subspace, otherwise $v(\boldsymbol x_i-\boldsymbol x_j)$ would not be preserved, in general. If the time reversal operation is such a block diagonal operator $Q$, 
it can be combined with the operator $P_{ij}$ that swaps particle $i$ with particle $j$, obtaining a new involution $QP_{ij}$: $QP_{ij}QP_{ij}=I$. The same applies to the case of half integer spin particles, and it is not necessary to repeat the reasoning.
Obviously, an interaction term, such as \eqref{poten}, prevents the use of the time reversal operations of Proposition \ref{prop:1.2.6}, that cannot be expressed by $3\times 3$ block diagonal matrices. Thus, in the following we neglect the case with $Q^2=-I$.

\end{rem}
\begin{rem}
\label{rem:2.1}
\normalfont
The action of the operator $\mathcal{T}$ of Proposition \ref{prop:2.3} deeply differs from that of the traditional one, which includes the inversion of the magnetic field, $\mathcal{T}_B$, cf.\ Ref.\cite{toda1991statistical}.
Nevertheless, $\mathcal{T}$ suffices for the validity of the Onsager relations, because they arise from a phase space integration and, pairing differently the contributions to the correlator, the integral may still yield the same result. 
Statistical mechanical relations are typically of this kind. Therefore, they do not require microreversibility. Weaker conditions suffice, such as those discussed here, that we may call \textit{statistical time reversibility}, cf.\ also Refs.\cite{Klages,ColR1,ColR2}.
This equally applies to the quantum case, if trajectories are replaced by the time evolution of states in the Hilbert space. In fact, the invariance of the Hamiltonian and of the equation of motion ensures that
$$ 
\ket{\psi'(t')}=\mathcal{T}\ket{\psi(t')}\, , \quad \mbox{with } ~~ t'=-t
$$
is a physical state if $\ket{\psi(t)}$ is. Roughly speaking, in the quantum case it is just a matter of rearranging the contributions to the trace in \eqref{eq:2.23}.
\end{rem}
Let us reconsider, in the light of the above result, the case of a constant magnetic field.
\begin{theo}
Take a system of particles interacting via 
the central potential \eqref{poten}. Let this system be subjected to a constant external magnetic field, $\boldsymbol B=(B_1,B_2,B_3)$. The dynamics is invariant under infinitely many time reversal operations: those whose action on each single particle subspace is represented by:
\begin{equation}
\mathcal{M}_m=
    \begin{pmatrix}
        a&b&0\\
        b&-a&0\\
        0&0& 1
    \end{pmatrix}
    \label{teosd}
\end{equation}
where $a,b \in \mathbb{R}$, $a^2+b^2=1$ and $\operatorname{det}(\mathcal{M}_m)=-1$.
\end{theo}
\begin{proof}
The time evolution of the system does not depend on the coordinates frame, so we can choose the axis $z$ along the direction of $\boldsymbol B$, and write
$B=(0,0,1)$, up to a dimensional constant. Consider the time reversal transformations whose action on each single particle subspace is represented by Eq.\eqref{teosd}.
By definition, $\mathcal{M}_m\in O(3)$ and $\mathcal{M}_m^2=I$, therefore $\mathcal{M}_m$ represents the action on coordinates of a time reversal operation, as follows from Theorem \ref{theo:1.1.2}. Furthermore, the compatibility condition of Theorem \ref{theo:3.1.2} is trivially verified:
\begin{equation}
    \operatorname{det}(\mathcal{M}_m) \mathcal{M}_m \boldsymbol B =(0,0,-1) = -\boldsymbol B
\end{equation}
Taking $a=\cos{\theta}$ and $b=\sin{\theta}$, with $\theta\in[0,2\pi)$, infinitely many possible choices are allowed for $\mathcal{M}_m$.
\end{proof}
This result seems to boldly contradict the traditional opinion that 
any magnetic field 
breaks TRI. 
With hindsight, it is not so surprising, when the existence of generalized time reversal symmetries has been ascertained.
In particular, a constant magnetic field directed along along the $z$ axis preserves all the symmetries that differ by rotations of the $xy$ plane, if it preserves one of them.

In the case of particles with spin, the Pauli Hamiltonian \eqref{eq:2.40} contains the minimal coupling term, for which 
Proposition \ref{prop:3.1.2} applies.
However, magnetic field and spins are also coupled, with the spins belonging to a different space. Then, a time reversal operation ought to take the form: 
\begin{equation}
    \mathcal{T}=(U_c\otimes U_s^1\otimes ...\otimes U_s^N)K
    \label{UcUs1UsN}
\end{equation}
where $U_c$ acts on coordinate and momentum operators, while $U_s^j$ acts on the spin operator of particle $j$. The situation differs from that considered in Theorem \ref{theo:2.7}, because we now have:
\begin{equation}
    \mathcal{T}H\mathcal{T}^{-1}\propto \boldsymbol B(U_c\boldsymbol x_j U_c^{-1})\cdot U_s^j K\boldsymbol \sigma_j K U_s^j
\end{equation}
where $H$ is the Hamiltonian and $\boldsymbol B$ the magnetic field. This way, the minimal coupling and the spin couplings with $\boldsymbol B$ are considered independently of each other:
a time reversal operation compatible with the minimal coupling leaves the spin field coupling unchanged. 
\begin{rem}
The transformation $U_c$ acts on coordinates and not on $\boldsymbol B$ as a vector field, that is the components of $\boldsymbol B$ transform as \begin{equation}
    U_c B_i U_c^{-1}\equiv B_i(U_c\boldsymbol x_i U_c^{-1})
\end{equation}
\end{rem}
\noindent
Note that statistical relations could be used to distinguish systems whose particles possess spin or from those which do not, if the extra conditions implied by the presence of spin reduced the number of suitable time reversal operators. In fact, for transformations like \eqref{UcUs1UsN}, and with block diagonal $U_c$ separately acting on single particles subspaces, this is not possible.
\begin{theo}
\label{theo:2.10}
The compatibility condition of Theorem \ref{theo:3.1.2} suffices for TRI to hold also in case of particles with spin that obey the Pauli equation \eqref{eq:2.40}.
\end{theo}

\begin{proof}
Write the Pauli hamiltonian as:
\begin{equation}
    H=H_{mc}+H_{sc}
\end{equation}
where $H_{mc}$ refers to the minimal coupling, of the spinless case, and $H_{sc}=\sum_{j=1}^N \boldsymbol{\sigma}_j\cdot \boldsymbol B(\boldsymbol x_j)$ is the spin-field coupling, with $\boldsymbol\sigma=(\sigma_x,\sigma_y,\sigma_z)$. Given that we are analyzing transformations separately acting on single particle spaces, we study a single addend of the summation. 

By hypothesis, there exists a time reversal operator $\mathcal{T}=(U_c\otimes U_s)K$ whose action on coordinates commutes with $H_{mc}$, \textit{i.e.} Eq.\eqref{3.22} holds or, equivalently,
\begin{equation}
\label{2.136}
   \boldsymbol{B}(\mathcal{M}_m\boldsymbol x)=-\frac{1} {\operatorname{det}(\mathcal{M}_m)}\mathcal{M}_m^{-1}\boldsymbol{B}(\boldsymbol x)
\end{equation}
Because of Theorem \ref{theo:3.1.2}, 
$\mathcal{M}_m$ is the 3-dimensional orthogonal and involutory matrix representation of the action of the time reversal operator on a single particle position operator. Under the action of $U_c$, one has:
\begin{equation}
    U_c(\boldsymbol\sigma\cdot B(\boldsymbol x))U_c^{-1}=\boldsymbol\sigma\cdot (-P\boldsymbol B(\boldsymbol x))=-P\boldsymbol \sigma\cdot \boldsymbol B(\boldsymbol x)
\end{equation}
where $P\boldsymbol\sigma=P^k_j\sigma_k$, and 
\begin{equation}
    P=\frac{1}{\operatorname{det}(\mathcal{M}_m)}\mathcal{M}_m^{-1} 
\end{equation}
which is a special orthogonal matrix. By Theorem \ref{iso}, there exists a special unitary matrix $U$ such that 
\begin{equation}
    U \sigma_j U^{-1}=P^k_j\sigma_k
\end{equation}
Then, because $P^2=I$ by definition of time reversal operation, we have $U^2=\pm I$, since the isomorphism maps $\pm I\in SU(2)$ to $I\in SO(3)$. Moreover, letting $K$ be the complex conjugation operator, we have 
\begin{equation}
    \sigma_y K \sigma_j K\sigma_y=-\sigma_j
\end{equation}
where $j,k=1,2,3=x,y,z$. Now, take $U_s=U\sigma_y$. By definition, the time reversal operator acts on spin operators as
\begin{equation}
    \mathcal{T}\sigma_j\mathcal{T}^{-1}=U\sigma_y K \sigma_j K\sigma_y U^{-1}=-P_j^k\sigma_k
\end{equation}
where we used the decomposition of the Hilbert space of the system as the direct product of the spin space and of the coordinate space. This allows us to separately change the coordinate system in each space, and to write:
\begin{equation}
    \sum_{k=1}^N \boldsymbol{\sigma}_k\cdot \boldsymbol B(\boldsymbol x_k)=\sum_{k=1}^N 
    \Big[
    \boldsymbol{\sigma}_x\otimes\boldsymbol B_1(\boldsymbol x_k)+\boldsymbol{\sigma}_y\otimes\boldsymbol B_2(\boldsymbol x_k)+\boldsymbol{\sigma}_z\otimes\boldsymbol B_3(\boldsymbol x_k)
    \Big]
\end{equation}
As the choice of the $z$ axis in the spin space is not related to that of the $z$ axis in coordinate space, $U_s$ does not depend on the choice of $U_c$. This closes the circle: for any valid $U_c$, we can always find $U$, and so $U_s$, in order to keep $H_{sc}$ invariant. In particular, we obtain:
\begin{equation}
    \mathcal{T}\boldsymbol\sigma\cdot B(\boldsymbol x)\mathcal{T}^{-1}=-P\boldsymbol \sigma \cdot -P \boldsymbol B(\boldsymbol x)=\boldsymbol \sigma\cdot\boldsymbol B(\boldsymbol x)
\end{equation}
\end{proof}

\begin{lemma}
\label{lem1}
The Hamiltonian is not invariant under the action of $U_s=\sigma_y$ if the magnetic field is constant. To be recover the original Hamiltonian, the magnetic field must be manually inverted. 
\end{lemma}
\begin{proof}
The choice $U_s=\sigma_y$ corresponds to $U=I$, that is $P=I$ and $\mathcal{M}_m=\pm I$. This yields to two constraints on the magnetic field: $\boldsymbol B(\boldsymbol x)=-\boldsymbol B(\boldsymbol x)$ choosing the plus and $\boldsymbol B(-\boldsymbol x)=-\boldsymbol B(\boldsymbol x)$ with the minus, both absurd if $\boldsymbol B(\boldsymbol x)=\boldsymbol B$ is constant. But inverting $\boldsymbol B$ obviously restores the original Hamiltonian.
\end{proof}
\begin{lemma}
The operator $U_s=U\sigma_y$ is such that 
\begin{equation}
    U_s K U_s K=\pm I
\end{equation}
\end{lemma}
\begin{proof}
By direct computation 
\begin{equation}
    U_s K U_s K=U\sigma_y K U\sigma_y K= U \sigma_y U^{*}\sigma_y^{*} =- U \sigma_y U^{*}\sigma_y=-U^2
\end{equation}
where we used \eqref{complex}. This complete the proof. 
\end{proof}
\noindent
In particular, taking $U=I$ in order to lie in the case of Lemma \ref{lem1}, we find the usual relation $\sigma_y K\sigma_y K=-I$.

This means
that a class of time reversal operations, 
the ones separately acting on the single particle subspaces, can \textit{always} be absorbed by an internal change of coordinates in spin space. Consequently, one finds that, \textit{contrary} to what is commonly believed, the presence of spin does not prevent 
time reversibility.


\subsection{Application of generalized TRI}
\label{sec:example}
Let us consider the calculation of Ref.\cite{bonella2017time}, concerning the diffusion tensor for a particle system in the presence of a constant magnetic field along the $z$ axis:
\begin{equation}
\label{3.54}
    D_{\alpha\beta}\propto\langle v_i^\alpha(0) v_j^\beta (t)\rangle\;\;\;\; \forall i,j
    \, ; \quad \alpha,\beta=x,y,z
\end{equation}
where $\alpha,\beta=x,y,z$ and $i,j$ are  particle labels. In particular, Ref.\cite{rondom} proved that the
correlator of velocities \eqref{3.54}
vanishes if there are two time reversal operations that map $p_j^\beta(t)$ respectively in $p_k^\gamma(-t)$ and $-p_k^\gamma(-t)$ (the case with $i=j$ and $\alpha=\beta$ simultaneously being excluded), while they act in the same way on $p_i^\alpha(t)$.
The proof simply observes that two expressions must be satisfied at once:
\begin{equation}
    \begin{cases}
    \langle v_i^\alpha(0) v_j^\beta (t)\rangle=(\pm 1)(+1)\langle v_i^\alpha(0) v_k^\gamma (-t)\rangle\\
    \langle v_i^\alpha(0) v_j^\beta (t)\rangle=(\pm 1)(-1)\langle v_i^\alpha(0) v_k^\gamma(-t)\rangle
    \end{cases}
\end{equation}
which only happens if the correlator vanishes.
The case $i=j$ and $\alpha=\beta$ is not included, since the autocorrelation $\langle v_i^\alpha(0) v_i^\alpha (t)\rangle$ is always mapped in another autocorrelation with plus sign. This example shows how 
practical the apparently abstract notion of TRI can be. In the present case, it allows a direct evaluation of transport coefficients.

For the classical case, we may now extend the result of \cite{bonella2017time} considering a more general form of magnetic field.
\begin{prop}
Let a system of coupled particles be subjected to an external magnetic field directed along the $z$-axis, $\boldsymbol B=B(x,y)\hat{\boldsymbol k}$. Assume $B(x,y)=B(x,-y)$ and $B(x,y)=B(y,x)$. Then, the diffusion tensor obeys:
\begin{equation}
    D_{xy}=-D_{yx}
\end{equation}
\label{Prot29}
\end{prop}
\begin{proof}
Notice first that Eq.\eqref{3.22} is verified by a magnetic field like the one in this Proposition. For example, consider the time reversal operation on the single particle subspace defined as
\begin{equation}
    \mathcal{M}(x,y,z,p_x,p_y,p_z)=(y,x,z,p_y,p_x,p_z)
\end{equation}
which implies 
\begin{equation}
    \mathcal{M}_m=
    \begin{pmatrix}
        0&1&0\\
        1&0&0\\
        0&0&1
    \end{pmatrix}
\end{equation}
with the notation of Theorem \ref{theo:3.1.2}.
Equation \eqref{3.22} then holds for such a transformation and the magnetic field.
Applying the same to all the particle subspaces and computing the correlators we obtain
\begin{equation}
    \langle v_i^x(0) v_j^y (t)\rangle=\langle v_i^y(0) v_j^x (-t)\rangle=-\langle v_i^y(0) v_j^x (t)\rangle
\end{equation}
that means $D_{xy}=-D_{yx}$.
\end{proof}

\noindent
Examples of magnetic fields that satisfy  Proposition \ref{Prot29} are $B(x,y)=const$ and $B(x,y)=B(x^2+y^2)$, respectively the case of a constant magnetic field and of a magnetic field depending on the distance from the $z$ axis. The result cannot be obtained in term of the diagonal operations only, because the diagonal operations cannot disentangle the pairs of subscripts and superscripts $(i,x)$ and $(j,y)$. 


\section{\large Results and Discussion}
In this paper we extended the list of generalized time reversal transformations, to include operators that swap the coordinates of different particles. This is allowed by the fact that a point in a $6N$-dimensional space does not distinguish the nature of its single components. The formal definition of TRI does not prevent the use of this kind of operations. Their importance arises in particular when one studies systems of particles with the same mass and charge. These are important in the present context, since investigations of the Onsager relations rarely deal with more than two species of particles.

Moreover, we extended this treatment to the quantum mechanical framework, in the wake of the work of Ref.\cite{rondoquantu}, about the $8$ diagonal time reversal operations on the single particle subspace. In particular, we investigated the swap operation in a context of non-relativistic Quantum Mechanics. 

We then defined generalized TRI for systems of particles with spin $1/2$, described by the Pauli equation. In previous works, this was considered out of reach \cite{rondoquantu}, because spin was believed to irremediably break TRI. On the contrary, we found combined sufficient conditions, concerning the generalized time reversal transformations and the form of the magnetic field, for the validity of TRI. That allow us to derive Onsager relations, as well as other relations requiring TRI, to hold in quantum systems coupled with an external magnetic field. 

This is interesting not only because of the experimental and theoretical relevance of such statistical mechanical relations, but also for the method used: we took full advantage of the fact that relations such as Onsager's are statistical relations. This allows many different paths to the same result and, in particular, microreversibility results unnecessary for Onsager and Fluctuation Relations.

We then developed an application to the calculation of the diffusion tensor, which uses different time reversal symmetries to conclude that certain correlators identically vanish.

The investigation, however, is not over. As recalled in Ref.\cite{casa}, violations of Onsager relations, which would imply the existence non-dissipative currents, have never been observed. And indeed Ref.\cite{casa} proved that those relations hold in cases in which one would have expected they are violated. At the same time, or theory does not exhaust all possible cases. 



Our results may also also pave the way to a new understanding of symmetries in quantum systems. The consequences of the generalized TRI have indeed rarely been fully investigated.
\vspace{1cm}\\
\small
\textbf{Conflict of interest}: The authors declare no conflict of interest.
\\
\textbf{Funding}: This research has been partially supported by Ministero dell’Istruzione, dell’Università e della Ricerca (MIUR) Grant No. E11G18000350001 “Dipartimenti di Eccellenza 2018-2022”.\\
\textbf{Aknowledgements}: This work has been performed under the auspices of Italian National Group of Mathematical
Physics (GNFM) of INdAM.

\printbibliography

\end{document}